\documentclass[AMA,STIX1COL]{WileyNJD-v2}
\usepackage{dsfont}
\usepackage{enumitem}
\usepackage[utf8]{inputenc}
\DeclareMathOperator{\logit}{logit}
\DeclareMathOperator{\expit}{expit}

\articletype{Research Article}%

\received{26 April 2016}
\revised{6 June 2016}
\accepted{6 June 2016}

\begin{document}

\title{Parametric G-computation for Compatible Indirect Treatment Comparisons with Limited Individual Patient Data}

\author[1,2]{Antonio Remiro-Az\'ocar}

\author[1,3,4]{Anna Heath}

\author[1]{Gianluca Baio}

\authormark{REMIRO-AZ\'OCAR \textsc{et al}}

\address[1]{\orgdiv{Department of Statistical Science}, \orgname{University College London}, \orgaddress{\state{London}, \country{United Kingdom}}}

\address[2]{\orgdiv{Quantitative Research}, \orgname{Statistical Outcomes Research \& Analytics (SORA) Ltd}, \orgaddress{\state{London}, \country{United Kingdom}}}

\address[3]{\orgdiv{Child Health Evaluative Sciences}, \orgname{The Hospital for Sick Children}, \orgaddress{\state{Toronto}, \country{Canada}}}

\address[4]{\orgdiv{Dalla Lana School of Public Health}, \orgname{University of Toronto}, \orgaddress{\state{Toronto}, \country{Canada}}}

\corres{*Antonio Remiro Az\'ocar, Department of Statistical Science, University College London, London, United Kingdom. \email{antonio.remiro.16@ucl.ac.uk}. Tel: (+44 20) 7679 1872. Fax: (+44 20) 3108 3105}

\presentaddress{Antonio Remiro Az\'ocar, Department of Statistical Science, University College London, Gower Street, London, WC1E 6BT, United Kingdom}

\abstract{Population adjustment methods such as matching-adjusted indirect comparison (MAIC) are increasingly used to compare marginal treatment effects when there are cross-trial differences in effect modifiers and limited patient-level data. MAIC is based on propensity score weighting, which is sensitive to poor covariate overlap and cannot extrapolate beyond the observed covariate space. Current outcome regression-based alternatives can extrapolate but target a conditional treatment effect that is incompatible in the indirect comparison. When adjusting for covariates, one must integrate or average the conditional estimate over the relevant population to recover a compatible marginal treatment effect. We propose a marginalization method based on parametric G-computation that can be easily applied where the outcome regression is a generalized linear model or a Cox model. The approach views the covariate adjustment regression as a nuisance model and separates its estimation from the evaluation of the marginal treatment effect of interest. The method can accommodate a Bayesian statistical framework, which naturally integrates the analysis into a probabilistic framework. A simulation study provides proof-of-principle and benchmarks the method's performance against MAIC and the conventional outcome regression. Parametric G-computation achieves more precise and more accurate estimates than MAIC, particularly when covariate overlap is poor, and yields unbiased marginal treatment effect estimates under no failures of assumptions. Furthermore, the marginalized regression-adjusted estimates provide greater precision and accuracy than the conditional estimates produced by the conventional outcome regression, which are systematically biased because the measure of effect is non-collapsible.}


\maketitle

\renewcommand{\thefootnote}{\alph{footnote}}

\section{Introduction}

The development of novel pharmaceuticals requires several stages, which include regulatory evaluation and, in several jurisdictions, health technology assessment (HTA).\cite{vreman2020decision} To obtain regulatory approval, a new technology must demonstrate efficacy. Well-conducted randomized controlled trials (RCTs) are the gold standard design for this purpose due to their internal validity, i.e., their potential for limiting bias within the study sample.\cite{temple2000placebo} Evidence supporting regulatory approval is often provided by a two-arm RCT, typically comparing the new technology to placebo or standard of care. Then, in certain jurisdictions, HTA addresses whether the health care technology should be publicly funded by the health care system. For HTA, manufacturers must convince payers that their product offers the best ``value for money'' of all available options in the market. This demands more than a demonstration of efficacy\cite{paul2001fourth} and will often require the comparison of treatments that have not been trialed against each other.\cite{sutton2008use}

In the absence of head-to-head trials, indirect treatment comparisons (ITCs) are at the top of the hierarchy of evidence to inform treatment and reimbursement decisions, and are very prevalent in HTA.\cite{dias2013evidence} Standard ITCs use indirect evidence obtained from RCTs through a common comparator arm.\cite{dias2013evidence, bucher1997results} These techniques are compatible with both individual patient data (IPD) and aggregate-level data (ALD). However, they are biased when the distribution of effect measure modifiers differs across trial populations, meaning that relative treatment effects are not constant.\cite{phillippo2018methods} 

% Standard propensity score methods\cite{austin2011introduction} can adjust for these differences in a pairwise comparison but require patient-level data for all studies.\cite{faria2015nice} 

Often in HTA, there are: (1) no head-to-head trials comparing the interventions of interest; (2) IPD available from the manufacturer's own trial but only published ALD for the comparator(s); and (3) imbalances in effect measure modifiers across studies. Several ``pairwise'' methods, labeled \textit{population-adjusted indirect comparisons}, have been introduced to estimate relative treatment effects in this scenario. These include matching-adjusted indirect comparison (MAIC),\cite{signorovitch2010comparative} based on inverse propensity score weighting, and simulated treatment comparison (STC),\cite{caro2010no} based on outcome modeling/regression. There is a simpler alternative, crude direct post-stratification,\cite{miettinen1972standardization} but this fails if any of the covariates are continuous or where there are several covariates for which one must account.\cite{stuart2011use} Very recently, a relevant outcome modeling-based approach called multilevel network meta-regression (ML-NMR) has been developed.\cite{phillippo2020multilevel, phillippo2019calibration} This incorporates larger networks of treatments and studies. The focus of this article is on the pairwise approaches but ML-NMR is considered in the discussion. 

Recommendations on the use of MAIC and STC in HTA have been provided, defining the relevant terminology and evaluating the theoretical validity of these methods.\cite{phillippo2018methods, phillippo2016nice} However, further research must: (1) examine these methods through comprehensive simulation studies; and (2) develop novel methods for population adjustment.\cite{phillippo2018methods, phillippo2016nice} In addition, recommendations have highlighted the importance of embedding the methods within a Bayesian framework, which allows for the principled propagation of uncertainty to the wider health economic model,\cite{baio2012bayesian} and is particularly appealing for ``probabilistic sensitivity analysis''.\cite{claxton2005probabilistic} This is a required component in the normative framework of HTA bodies such as NICE,\cite{baio2012bayesian, claxton2005probabilistic} used to characterize the impact of the uncertainty in the model inputs on decision-making.

Recently, several simulation studies have been conducted to assess population-adjusted indirect comparisons.\cite{remiro2020methods, cheng2019statistical, hatswell2020effects, phillippo2020assessing} Remiro-Az\'ocar et al. perform a simulation study benchmarking the performance of the typical use of MAIC and STC against the standard ITC for the Cox model and survival outcomes.\cite{remiro2020methods} In this study, MAIC yields unbiased and relatively accurate treatment effect estimates under no failures of assumptions, but the robust sandwich variance may underestimate standard errors where effective sample sizes are small. In the simulation scenarios, there is some degree of overlap between the studies' covariate distributions. Nevertheless, it is well known that weighting methods like MAIC are highly sensitive to poor overlap, are not asymptotically efficient, and incapable of extrapolation.\cite{phillippo2020assessing, stuart2010matching} With poor overlap, extreme weights may produce unstable treatment effect estimates with high variance. A related problem in finite samples is that feasible weighting solutions may not exist\cite{jackson2020alternative} due to separation problems where samples sizes are small and the number of covariates is large.\cite{van2011targeted, neugebauer2005prefer}

Outcome regression approaches such as STC are appealing as these tend to be more efficient than weighting, providing more stable estimators and allowing for model extrapolation.\cite{robins1992estimating, vo2021assessing}. We view extrapolation as an advantage because poor overlap, with small effective sample sizes and large percentage reductions in effective sample size, is a pervasive issue in HTA.\cite{phillippo2019population} While extrapolation can also be viewed as a disadvantage if it is not valid, in our case it expands the range of scenarios in which population adjustment can be used.

The aforementioned simulation study\cite{remiro2020methods} demonstrates that the typical usage of STC, as described by HTA guidance and recommendations,\cite{phillippo2016nice} produces systematically biased estimates of the marginal treatment effect, with inappropriate coverage rates, because it targets a conditional estimand instead. With the Cox model and survival outcomes, there is bias because the conditional (log) hazard ratio is non-collapsible. In addition, the conditional estimand cannot be combined in any indirect treatment comparison or compared between studies because non-collapsible conditional estimands vary across different covariate adjustment sets. This is a recurring problem in meta-analysis.\cite{hauck1998should, daniel2020making} 

The crucial element that has been missing from the typical usage of STC is the marginalization of treatment effect estimates. When adjusting for covariates, one must integrate or average the conditional estimate over the joint covariate distribution to recover a marginal treatment effect that is compatible in the indirect comparison. We propose a simple marginalization method based on parametric G-computation\cite{robins1986new, robins1987graphical} or model-based standardization,\cite{moore2009covariate, austin2010absolute} often applied in observational studies in epidemiology and medical research where treatment assignment is non-random. In meta-analysis, Vo et al.\cite{vo2021assessing,vo2019novel} have used parametric G-computation to transport RCT results to a specific target population. Our proposal extends these approaches to population-adjusted indirect comparisons with limited patient-level data. 

Parametric G-computation can be viewed as an extension to the conventional STC, making use of effectively the same outcome model. It is an outcome regression approach, thereby capable of extrapolation, that targets a marginal treatment effect. It does so by separating the covariate adjustment regression model from the evaluation of the marginal treatment effect of interest. The conditional parameters of the regression are viewed as nuisance parameters, not directly relevant to the research question. The method can be implemented in a Bayesian statistical framework, which explicitly accounts for relevant sources of uncertainty, allows for the incorporation of prior evidence (e.g.~expert opinion), and naturally integrates the analysis into a probabilistic framework, typically required for HTA.\cite{baio2012bayesian} 

In this paper, we carry out a simulation study to benchmark the performance of different versions of parametric G-computation against MAIC and the conventional STC. The simulations provide proof-of-principle and are based on scenarios with binary outcomes and continuous covariates, with the log-odds ratio as the measure of effect. 
% The methods are evaluated in 162 scenarios that vary the trial sample size, effect-modifying strength of covariates, prognostic effect of covariates, covariate overlap/imbalance and the level of correlation in the covariates. 
The parametric G-computation approaches achieve greater precision and accuracy than MAIC and are unbiased under no failures of assumptions. Furthermore, their marginal estimates provide greater precision than the conditional estimates produced by the conventional version of STC. While this precision comparison is irrelevant, because it is made for estimators of different estimands, it supports previous research on non-collapsible measures of effect.\cite{daniel2020making, moore2009covariate}  

In Section \ref{sec2}, we present the context and data requirements for population-adjusted indirect comparisons. Section \ref{sec3} provides a detailed description of the outcome regression methodologies. Section \ref{sec4} outlines a simulation study, which evaluates the statistical properties of different approaches to outcome regression with respect to MAIC. Section \ref{sec5} describes the results from the simulation study. An extended discussion of our findings is presented in Section \ref{sec6}. 

\section{Context}\label{sec2}

Consider an active treatment $A$, which needs to be compared to another active treatment $B$ for the purposes of reimbursement. Treatment $A$ is new and being tested for cost-effectiveness, while treatment $B$ is typically an established intervention, already on the market. Both treatments have been evaluated in a RCT against a common comparator $C$, e.g.~standard of care or placebo, but not against each other. Indirect comparisons are performed to estimate the relative treatment effect for a specific outcome. The objective is to perform the analysis that would be conducted in a hypothetical head-to-head RCT between $A$ and $B$, which indirect treatment comparisons seek to emulate. 

% The RCT is widely considered the gold standard design to evaluate treatments\cite{glenny2005indirect} due to its internal validity.\cite{temple2000placebo} Appropriate randomization guarantees covariate balance on expectation, so that the treatment groups are comparable and confounding is limited. Therefore, assuming no structural issues (e.g.~no dropout, non-adherence, measurement error, etc.), RCTs allow for unbiased estimation of the relative treatment effect within the study. 

RCTs have different types of potential target average estimands of interest: \textit{marginal} or \textit{population-average} effects, calibrated at the population level, and \textit{conditional} effects, calibrated at the individual level. The former are typically, but not necessarily, estimated by an ``unadjusted'' analysis. This may be a simple comparison of the expected outcomes for each group or a univariable regression including only the main treatment effect. Conditional treatment effects are typically estimated by an ``adjusted'' analysis (e.g.~a multivariable regression of outcome on treatment and covariates), accounting for prognostic variables that are pre-specified in the protocol or analysis plan, such as prior medical/treatment history, demographics and physiological status. A recurring theme throughout this article is that the terms ``conditional and adjusted (likewise marginal and unadjusted) should not be used interchangeably'' because marginal need not mean unadjusted and covariate-adjusted analyses may also target marginal estimands.\cite{daniel2020making, remiro2021target}

The marginal effect would be the average effect, at the population level (conditional on the entire population distribution of covariates), of moving all individuals in the trial population between two hypothetical worlds: from one where everyone receives treatment $B$ to one where everyone receives treatment $A$.\cite{austin2011introduction, hernan2020causal} The conditional effect corresponds to the average treatment effect at the unit level, conditional on the effects of the covariates that have also been included in the model. This would be the average effect of switching the treatment of an individual in the trial population from $B$ to $A$, fully conditioned on the average combination of subject-level covariates, or the average effect across sub-populations of subjects who share the same covariates. Population-adjusted indirect comparisons are used to inform reimbursement decisions in HTA at the population level. Therefore, marginal treatment effect estimates are required.\cite{remiro2021marginalization}

The indirect comparison between treatments $A$ and $B$ is typically carried out in the ``linear predictor'' scale;\cite{dias2013evidence, bucher1997results} namely, using additive effects for a given linear predictor, e.g.~log-odds ratio for binary outcomes or log hazard ratio for survival outcomes. Indirect treatment comparisons can be ``anchored'' or ``unanchored''. Anchored comparisons make use of a connected treatment network. In this case, this is available through a common comparator $C$. Unanchored comparisons use disconnected treatment networks or single-arm trials and require much stronger assumptions than their anchored counterparts.\cite{phillippo2018methods} The use of unanchored comparisons where there is connected evidence is discouraged and often labeled as problematic.\cite{phillippo2018methods, phillippo2016nice} This is because it does not respect within-study randomization and is not protected from imbalances in any covariates that are prognostic of outcome (almost invariably, a larger set of covariates than the set of effect measure modifiers). Hence, our focus is on anchored comparisons. 

In the standard anchored scenario, a manufacturer submitting evidence to HTA bodies has access to IPD from its own trial that compares its treatment $A$ against the standard health technology $C$. The disclosure of proprietary, confidential IPD from industry-sponsored clinical trials is rare. Hence, individual-level data on baseline covariates, treatment and outcomes for the competitor's trial, evaluating the relative efficacy or effectiveness of intervention $B$ vs.~$C$, are regularly unavailable, for both the submitting company and the national HTA agency assessing the evidence. For this study, only summary outcome measures and marginal moments of the covariates, e.g.~means with standard deviations for continuous variables or proportions for binary and categorical variables, as found in a table of baseline characteristics in clinical trial publications, are available. We consider, without loss of generality, that IPD are available for a study comparing treatments $A$ and $C$ (denoted $AC$) and published ALD are available for a study comparing interventions $B$ and $C$ ($BC$).

Standard ITCs such as the Bucher method\cite{bucher1997results} assume that there are no differences across trials in \textit{effect measure modifiers}, \textit{effect modifiers} for short. The relative effect of a particular intervention, as measured on a specific scale, varies at different levels of the effect modifiers. Within the biostatistics literature, effect modification is usually referred to as heterogeneity or interaction, because effect modifiers are considered to alter the effect of treatment by interacting with it on a specific scale,\cite{vanderweele2009concerning} and are typically detected by examining statistical interactions.\cite{rothman1980concepts}

Consider that $Z$ denotes a treatment indicator. Active treatment $A$ is denoted $Z=1$, active treatment $B$ is denoted $Z=2$, and the common comparator $C$ is denoted $Z=0$. In addition, $S$ denotes a specific study. The $AC$ study, comparing treatments $A$ and $C$ is denoted $S=1$. The $BC$ study is denoted $S=2$. The true relative treatment effect between $Z$ and $Z'$ in study population $S$ is indicated by $\Delta_{ZZ'}^{(S)}$ and is estimated by $\hat{\Delta}_{ZZ'}^{(S)}$.

In standard ITCs, one assumes that the $A$ vs.~$C$ treatment effect $\Delta_{10}^{(1)}$ in the $AC$ population is equal to $\Delta_{10}^{(2)}$, the effect that would have have occurred in the $BC$ population. Note that the Bucher method and most conventional network meta-analysis methods do not explicitly specify a target population of policy interest (whether this is $AC$, $BC$ or otherwise).\cite{manski2019meta} Hence, they cannot account for differences across study populations. The Bucher method is only valid when either: (1) the $A$ vs.~$C$ treatment effect is homogeneous, such that there is no effect modification; or (2) the distributions of the effect modifiers are the same in both studies. 

If the $A$ vs.~$C$ treatment effect is heterogeneous and the effect modifiers are not equidistributed across trials, relative treatment effects are no longer constant across the trial populations, except in the pathological case where the bias induced by different effect modifiers is in opposite directions and cancels out. Hence, the assumptions of the Bucher method are broken. In this scenario, standard ITC methods are liable to produce biased and overprecise estimates of the treatment effect.\cite{song2003validity} These features are undesirable, particularly from the economic modeling point of view, as they impact negatively on the probabilistic sensitivity analysis.

Conversely, MAIC and STC target the $A$ vs.~$C$ treatment effect that would be observed in the $BC$ population, thereby performing an adjusted indirect comparison in such population. MAIC and STC implicitly assume that the target population is the $BC$ population. The estimate of the adjusted $A$ vs.~$B$ treatment effect is:
\begin{equation}
\hat{\Delta}_{12}^{(2)} = \hat{\Delta}_{10}^{(2)} - \hat{\Delta}_{20}^{(2)},
\label{eqn1}
\end{equation}
where $\hat{\Delta}_{10}^{(2)}$ is the estimated relative treatment effect of $A$ vs $C$ (mapped to the $BC$ population), and $\hat{\Delta}_{20}^{(2)}$ is the estimated marginal treatment effect of $B$ vs.~$C$ (in the $BC$ population). The estimate $\hat{\Delta}_{20}^{(2)}$ and an estimate of its variance may be directly published or derived non-parametrically from crude aggregate outcomes made available in the literature. The majority of RCT publications will report an estimate targeting a marginal treatment effect, derived from a simple regression of outcome on a single independent variable, treatment assignment. In addition, the estimate $\hat{\Delta}_{12}^{(2)}$ should target a marginal treatment effect for reimbursement decisions at the population level. Therefore, $\hat{\Delta}_{10}^{(2)}$ should target a marginal treatment effect that is compatible with $\hat{\Delta}_{20}^{(2)}$.\cite{remiro2020conflating}

As the relative effects, $\hat{\Delta}_{10}^{(2)}$ and $\hat{\Delta}_{20}^{(2)}$, are specific to separate studies, the within-trial randomization of the originally assigned patient groups is preserved. Because the estimates are based on different study samples (IPD are unavailable for $BC$), the within-trial relative effects are assumed statistically independent of each other. Hence, their variances are simply summed to estimate the variance of the $A$ vs.~$B$ treatment effect. One can also take a Bayesian approach to estimating the indirect treatment comparison, in which case variances would be derived empirically from draws of the posterior density. In our opinion, a Bayesian analysis is helpful because simulation from the posterior distribution provides a framework for probabilistic decision-making, directly allowing for both statistical estimation and inference, and for principled uncertainty propagation.\cite{dias2013evidence} 

A reference intervention is required to define the effect modifiers. In the methods considered in this article, we are selecting the effect modifiers of treatment $A$ with respect to $C$ (as opposed to the treatment effect modifiers of $B$ vs.~$C$). This is because we have to adjust for these in order to perform the indirect comparison in the $BC$ population, implicitly assumed to be the target population. If one had access to IPD for the $BC$ study and only published ALD for the $AC$ study, one would have to adjust for the factors modifying the effect of treatment $B$ with respect to $C$, in order to perform the comparison in the $AC$ population.

In some contexts, a distinction is made between sample-average and population-average marginal effects.\cite{stuart2011use, cole2010generalizing} In this article, ``population-adjusted'' indirect comparisons refer to ``sample-adjusted'' indirect comparisons because, due to patient-level data limitations, the methods contrast treatments in the $BC$ trial sample. Typically, an implicit assumption is that the sample on which inferences are made (as described by its published covariate summaries for $\hat{\Delta}_{10}^{(2)}$) is exactly the trial's target population. Alternatively, the assumption is that the study sample is a random sample, i.e., representative, of such population, ignoring sampling variability in the patients' baseline characteristics and assuming that no random error attributable to such exists. In reality, the subjects of the $BC$ study have been sampled from a, typically more diverse, target population of eligible patients, defined by the trial's inclusion and exclusion criteria. 

% \subsection{The need for outcome regression approaches}\label{subsec21}

\subsection{Some assumptions}\label{subsec22}

MAIC and the outcome regression methods discussed in this article mostly require the same set of assumptions. A non-technical description of these is detailed in Appendix B of the Supplementary Material, along with some discussion about potential failures of assumptions and their consequences in the context of the simulation study. The assumptions are: 
\begin{enumerate}
    \item Internal validity of the $AC$ and $BC$ trials, e.g.~appropriate randomization and sufficient sample sizes so that the treatment groups are comparable, no interference, negligible measurement error or missing data, the absence of non-compliance, etc.
    \item Consistency under parallel studies such that both trials have identical control treatments, sufficiently similar study designs and outcome measure definitions, and have been conducted in care settings with a high degree of similarity. 
    \item Accounting for all effect modifiers of treatment $A$ vs.~$C$ in the adjustment. This assumption is called the conditional constancy of the $A$ vs.~$C$ marginal treatment effect,\cite{phillippo2016nice} and requires that a sufficiently rich set of baseline covariates has been measured for the $AC$ study and is available in the $BC$ study publication.\footnote{In the anchored scenario, we are interested in a comparison of \textit{relative} outcomes or effects, not \textit{absolute} outcomes. Hence, an anchored comparison only requires conditioning on the effect modifiers of the $A$ vs.~$C$ treatment effect. This assumption is named the \textit{conditional constancy of relative effects},\cite{phillippo2018methods, phillippo2016nice} i.e., given the selected effect-modifying covariates, the marginal $A$ vs.~$C$ treatment effect is constant across the $AC$ and $BC$ populations. There are other formulations of this assumption,\cite{stuart2011use, cole2010generalizing} such as trial assignment/selection being conditionally ignorable, unconfounded or exchangeable for such treatment effect, i.e., conditionally independent of the treatment effect, given the selected effect modifiers. One can consider that being in population $AC$ or population $BC$ does not carry over any information about the marginal $A$ vs $C$ treatment effect, once we condition on the treatment effect modifiers. This means that after adjusting for these effect modifiers, treatment effect heterogeneity and trial assignment are conditionally independent. Another advantage of outcome regression with respect to weighting is that, by being less sensitive to overlap issues, it allows for the inclusion of larger numbers of effect modifiers. This makes it easier to satisfy the conditional constancy of relative effects.}
    \item Overlap between the covariate distributions in $AC$ and $BC$. More specifically, that the ranges of the selected covariates in the $AC$ trial cover their respective moments in the $BC$ population. The overlap assumption (often referred to as ``positivity'') can be overcome in outcome regression if one is willing to rely on model extrapolation, assuming correct model specification.\cite{hernan2020causal} 
    \item Correct specification of the $BC$ population. Namely, that it is appropriately represented by the information available to the analyst, that does not have access to patient-level data from the $BC$ study. As the full joint distribution of covariates is unavailable for $BC$, this population is characterized by a combination of the covariate moments published for the $BC$ study, and some assumptions about the covariates' correlation structure and marginal distribution forms. 
    \item Correct (typically parametric) model specification. This assumption is different for MAIC and outcome regression. In MAIC, a logistic regression is used to model the trial assignment odds conditional on a selected set of baseline covariates. The weights estimated by the model represent the ``trial selection'' odds, namely, the odds of being enrolled in the $BC$ trial.\footnote{Note that the``matching-adjusted'' term in MAIC is a misnomer, as the indirect comparison is actually ``weighting-adjusted''.} MAIC does not explicitly require an outcome model. On the other hand, outcome regression methods estimate an outcome-generating mechanism given treatment and the baseline covariates. While MAIC relies on a correctly specified model for the weights given the covariates, outcome regression methods rely on a correctly specified model for the conditional expectation of the outcome given treatment and the covariates.
    
    % \item A final assumption, the shared effect modifier assumption (described in subsection \ref{subsec43}), is required to transport the treatment effect estimate for $A$ vs.~$B$ from the $BC$ population to any other target population. Otherwise, one has to assume that the $BC$ population is the target for the analysis to be valid.
\end{enumerate}
Most assumptions are causal and untestable, with their justification typically requiring prior substantive knowledge.\cite{remiro2020principled} Nevertheless, we shall assume that they hold throughout the article. 

\section{Methodology}\label{sec3}

\subsection{Data structure}\label{subsec31}

For the $AC$ trial IPD, let $\mathcal{D}_{AC} = ({\boldsymbol{x}, \boldsymbol{z}, \boldsymbol{y}})$. Here, $\boldsymbol{x}$ is a matrix of baseline characteristics (covariates), e.g.~age, gender, comorbidities, baseline severity, of size $N \times K$, where $N$ is the number of subjects in the trial and $K$ is the number of available covariates. For each subject $n=1,\dots,N$, a row vector $\boldsymbol{x}_n$ of $K$ covariates is recorded. Each baseline characteristic can be classed as a prognostic variable (a covariate that affects outcome), an effect modifier, both or none. For simplicity in the notation, it is assumed that all available baseline characteristics are prognostic of the outcome and that a subset of these, $\boldsymbol{x}^{\boldsymbol{(EM)}} \subseteq \boldsymbol{x}$, is selected as effect modifiers on the linear predictor scale, with a row vector $\boldsymbol{x}^{(EM)}_n$ recorded for each subject. We let $\boldsymbol{y} = (y_1, y_2, \dots, y_N)$ represent a vector of outcomes, e.g.~a time-to-event or binary indicator for some clinical measurement; and $\boldsymbol{z} = (z_1, z_2, \dots, z_N)$ is a treatment indicator ($z_n=1$ if subject $n$ is under treatment $A$ and $z_n=0$ if under $C$). For simplicity, we shall assume that there are no missing values in $\mathcal{D}_{AC}$. The outcome regression methodologies can be readily adapted to address this issue, particularly under a Bayesian implementation, but this is an area for future research.

We let $\mathcal{D}_{BC}=[\boldsymbol{\theta}, \boldsymbol{\rho}, \hat{\Delta}_{20}^{(2)}, \hat{V}(\hat{\Delta}_{20}^{(2)})]$ denote the information available for the $BC$ study. No individual-level information on covariates, treatment or outcomes is available. Here, $\boldsymbol{\theta}$ represents a vector of published covariate summaries, e.g.~proportions or means. For ease of exposition, we shall assume that these are available for all $K$ covariates (otherwise, one would take the intersection of the available covariates), and that the selected effect modifiers are also available such that $\boldsymbol{\theta}^{\boldsymbol{(EM)}} \subseteq \boldsymbol{\theta}$. An estimate $\hat{\Delta}_{20}^{(2)}$ of the $B$ vs.~$C$ treatment effect in the $BC$ population, and an estimate of its variance $\hat{V}(\hat{\Delta}_{20}^{(2)})$, either published directly or derived from crude aggregate outcomes in the literature, are also available. Note that these are not used in the adjustment mechanism but are ultimately required to perform inference for the indirect comparison in the $BC$ population. 

Finally, we let the symbol $\boldsymbol{\rho}$ stand for the dependence structure of the $BC$ covariates. Under certain assumptions about representativeness, this can be retrieved from the $AC$ trial, e.g.~through the observed pairwise correlations, or from external data sources such as registries. This information, together with the published covariate summary statistics, is required to characterize the joint covariate distribution of the $BC$ population. A pseudo-population of $N^*$ subjects is simulated from this joint distribution, such that $\boldsymbol{x}^*$ denotes a matrix of baseline covariates of dimensions $N^* \times K$, with a row vector $\boldsymbol{x}_i^*$ of $K$ covariates simulated for each subject $i=1,\dots N^*$. Notice that the value of $N^*$ does not necessarily have to correspond to the actual sample size of the $BC$ study; however, the simulated cohort must be sufficiently large so that the sampling distribution is stabilized, minimizing sampling variability. Again, a subset of the simulated covariates, $\boldsymbol{x}^{*\boldsymbol{(EM)}} \subseteq \boldsymbol{x^*}$, makes up the treatment effect modifiers on the linear predictor scale, with a row vector $\boldsymbol{x}_i^{*\boldsymbol{(EM)}}$ for each subject $i=1,\dots,N^*$. In this article, the asterisk superscript represents unobserved quantities that have been constructed in the $BC$ population. 

The outcome regression approaches discussed in this article estimate treatment effects with respect to a hypothetical pseudo-population for the $BC$ study. Before outlining the specific outcome regression methods, we explain how to generate values for the individual-level covariates $\boldsymbol{x^*}$ for the $BC$ population using Monte Carlo simulation. 

\subsection{Individual-level covariate simulation}\label{subsec32}

% Ideally, the $BC$ population should be characterized by the full joint distribution of covariates. However, the restriction of limited IPD makes it unlikely that the joint distribution of the $BC$ covariates is available. Where there are not many covariates and these are binary, this is sometimes available as a cross-tabulation. However, most of the time we need to approximate the joint distribution appropriately. This is important to avoid bias arising from the incomplete specification of the $BC$ population. The published summary values $\boldsymbol{\theta}$ and the correlation structure $\boldsymbol{\rho}$ are combined, making certain parametric assumptions about the marginal distributional forms, to infer the joint distribution of the $BC$ covariates and construct an appropriate pseudo-population for inferences. The proposed approaches allow the analyst to bring in some prior knowledge or evidence to inform the potential distributions of the covariates. However, it is worth noting that we cannot give a general recipe for this step, which requires context-specific knowledge that is likely not available from the observed data in the trials. 

Firstly, the marginal distributions for each covariate are specified. The mean and, if applicable, the standard deviation of the marginals are sourced from the $BC$ report to match the published summary statistics. As the true marginal distributional forms are not known, some parametric assumptions are required. For instance, if it is reasonable to assume that sampling variability for a continuous covariate can be described using a normal distribution, and the covariate's mean and standard deviation are published in the $BC$ report, we can assume that it is marginally normally distributed. Hence, we can also select the family for the marginal distribution using the theoretical validity of the candidate distributions alongside the IPD. For example, the marginal distribution of duration of prior treatment at baseline could be modeled as a log-normal or Gamma distribution as these distributions are right-skewed and bounded to the left by zero. Truncated distributions can be used to resemble the inclusion/exclusion criteria for continuous covariates in the $BC$ trial, e.g.~age limits, and avoid deterministic overlap violations.  

Secondly, the correlations between covariates are specified. We suggest two possible data-generating model structures for this purpose: (1) simulating the covariates from a multivariate Gaussian copula;\cite{phillippo2020multilevel, nelsen2007introduction} or (2) factorizing the joint distribution of the covariates into the product of marginal and conditional distributions.
%; for instance, if the covariates are $\boldsymbol{x}=(x_1,x_2)$ indicating age and sex, respectively, we may write $p(x_1,x_2)=p(x_2)p(x_1\mid x_2)$, where the marginal distribution over sexes and the conditional distribution for age (groups) by sexes could be informed using census data. 
The former approach is perhaps more general-purpose. The latter is more flexible, defining separate models for each variable, but its specification can be daunting where there are many covariates and interdependencies are complex. 

Any multivariate joint distribution can be decomposed in terms of univariate marginal distribution functions and a dependence structure.\cite{sklar1959fonctions} A Gaussian copula ``couples'' the marginal distribution functions for each covariate to a multivariate Gaussian distribution function. The main appeal of a copula is that the correlation structure of the covariates and the marginal distribution for each covariate can be modeled separately. We may use the pairwise correlation structure observed in the $AC$ patient-level data as the dependence structure, while keeping the marginal distributions inferred from the $BC$ summary values and the IPD. Note that the term ``Gaussian'' does not refer to the marginal distributions of the covariates but to the correlation structure. While the Gaussian copula is sufficiently flexible for most modeling purposes, more complex copula types (e.g.~Clayton, Gumbel, Frank) may provide different and more customizable correlation structures.\cite{nelsen2007introduction}

\begin{figure}[!htb]
\center{\includegraphics[width=0.48\textwidth]{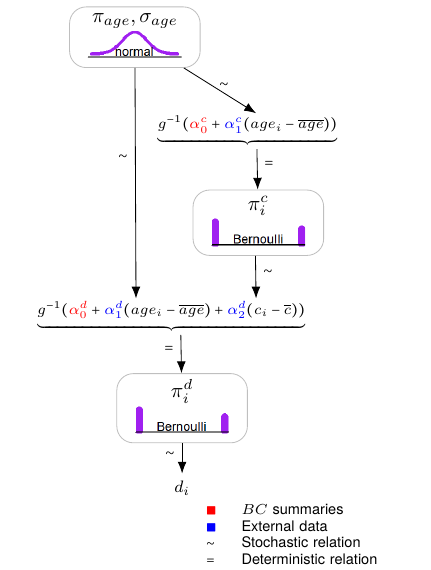}}
\caption[]
{
An example of individual-level Monte Carlo covariate simulation where the joint distribution of three baseline characteristics, $age$, comorbidity $c$ and comorbidity $d$, is factorized into the product of marginal and conditional distributions, such that $p(age, c, d)=p(d \mid c, age)p(c \mid age)p(age)$. The joint distribution is valid because the conditional distributions defining the covariates are compatible: we start with a marginal distribution for age and construct the joint distribution by modeling each additional covariate, one-by-one, conditionally on the covariates that have already been simulated. This diagram adopts the convention of Kruschke.\cite{kruschke2014doing}}
\label{fig1}
\end{figure}

\clearpage

Alternatively, we can account for the correlations by factorizing the joint distribution of covariates in terms of marginal and conditional densities. This strategy is common in implementations of sequential conditional algorithms for parametric multiple imputation.\cite{royston2004multiple, buuren2010mice} For instance, consider two baseline characteristics: $age$, which is a continuous variable, and the presence of a comorbidity $c$, which is dichotomous. We can factorize the joint distribution of the covariates such that $p(age, c)=p(c \mid age)p(age)$. 

In this scenario, we draw $age_i$ for subject $i$ from a suitable marginal distribution, e.g.~a normal, with the mean and standard deviation sourced from the published $BC$ summaries or official life tables. The mean $\pi_i^{c}$ of $c$ (the conditional proportion of the comorbidity) given the age, can be modeled through a regression: $\pi_i^{c} = g^{-1}(\alpha_0^{c} + \alpha_1^{c}(age_i - \overline{age}))$, with $c_i \sim \textnormal{Bernoulli}(\pi_i^c)$ where $g(\cdot)$ is an appropriate link function. Here, the coefficients $\alpha_0^{c}$ and $\alpha_1^{c}$ represent respectively the overall proportion of comorbidity $c$ in the $BC$ population (marginalizing out the age), and the correlation level between comorbidity $c$ and (the centered version of) age. The former coefficient can be directly sourced from the published $BC$ summaries, whereas the latter could be derived from pairwise correlations observed in the $AC$ IPD or from external sources, e.g.~clinical expert opinion, registries or administrative data, applying the selection criteria of the $BC$ trial to subset the data. Figure \ref{fig1} provides an example of a similar probabilistic structure with three covariates: $age$ and the presence of two comorbidities, $c$ and $d$. In this example, the distribution of the covariates is factorized such that $p(age, c, d)=p(d \mid c, age)p(c \mid age)p(age)$. 

\subsection{Conventional outcome regression}\label{subsec33}

In simulated treatment comparison (STC)\cite{caro2010no}, IPD from the $AC$ trial are used to fit a regression model describing the observed outcomes $\boldsymbol{y}$ in terms of the relevant baseline characteristics $\boldsymbol{x}$ and the treatment variable $\boldsymbol{z}$. 

STC has different formulations.\cite{phillippo2018methods, caro2010no, phillippo2016nice, ishak2015simulation} In the conventional version described by the NICE Decision Support Unit Technical Support Document 18,\cite{phillippo2018methods, phillippo2016nice} the individual-level covariate simulation step described in subsection \ref{subsec32} is not performed. The covariates are centered at the published mean values $\boldsymbol{\theta}$ from the $BC$ population. Under a generalized linear modeling framework, the following regression model is fitted to the observed $AC$ IPD:
\begin{equation}
g(\mu_n) = \beta_0 +  \left(\boldsymbol{x}_n  - \boldsymbol{\theta}  \right)\boldsymbol{\beta_1} + \left[\beta_z +  \left(\boldsymbol{x}_n^{\boldsymbol{(EM)}} - \boldsymbol{\theta}^{\boldsymbol{(EM)}} \right) \boldsymbol{\beta_2}\right]\mathds{1}(z_n=1),
\label{eqn2}
\end{equation}
where, for a generic subject $n$, $\mu_n$ is the expected outcome on the natural scale, e.g.~the probability scale for binary outcomes, $g(\cdot)$ is an invertible canonical link function, $\beta_0$ is the intercept, $\boldsymbol{\beta_1}$ is a vector of $K$ regression coefficients for the prognostic variables, $\boldsymbol{\beta_2}$ is a vector of interaction coefficients for the effect modifiers (modifying the $A$ vs.~$C$ treatment effect) and $\beta_z$ is the $A$ vs.~$C$ treatment coefficient. In the population adjustment literature, covariates are sometimes centered separately for active treatment and common comparator arms.\cite{petto2019alternative, belger2015inclusion} We do not recommend this approach because it is likely to break randomization, distorting the balance between treatment arms $A$ and $C$ on covariates that are not accounted for. If these covariates are prognostic of outcome, this would compromise the internal validity of the within-study treatment effect estimate for $A$ vs.~$C$.

% For binary outcomes in logistic regression, one uses the $\logit(\mu_n) = \ln[\mu_n/(1-\mu_n)]$ link function, but other choices are possible in practice, e.g.~the identity link for standard linear regression with continuous-valued outcomes, or the log link for Poisson regression with count outcomes.  

The regression in Equation \ref{eqn2} models the conditional outcome mean given treatment and the centered baseline covariates. Because the IPD prognostic variables and the effect modifiers are centered at the published mean values from the $BC$ population, the estimated $\hat{\beta}_z$ is directly interpreted as the $A$ vs.~$C$ treatment effect in the $BC$ population or, more specifically, in a pseudopopulation with the $BC$ covariate means and the $AC$ correlation structure. Typically, analysts set $\hat{\Delta}_{10}^{(2)}=\hat{\beta}_z$ in Equation \ref{eqn1}, inputting this coefficient into the health economic decision model.\cite{grimm2019nivolumab, ren2019pembrolizumab} For uncertainty quantification purposes, the variance of said treatment effect is obtained from the standard error estimate of the treatment coefficient in the fitted model.\cite{phillippo2018methods, phillippo2016nice}

An important issue with this approach is that the treatment parameter $\hat{\beta}_z$, extracted from the fitted model, has a conditional interpretation at the individual level, because it is conditional on the baseline covariates included as predictors in the multivariable regression.\cite{remiro2020methods, austin2011introduction} However, we require that $\hat{\Delta}_{10}^{(2)}$ (and, subsequently, $\hat{\Delta}_{12}^{(2)}$) estimate a marginal treatment effect, for reimbursement decisions at the population level. In addition, we require that $\hat{\Delta}_{10}^{(2)}$ is compatible with the published marginal effect for $B$ vs.~$C$, $\hat{\Delta}_{20}^{(2)}$, for comparability in the indirect treatment comparison in Equation \ref{eqn1}. Even if the published estimate $\hat{\Delta}_{20}^{(2)}$ targets a conditional estimand, this cannot be combined in the indirect treatment comparison because it likely has been adjusted using a different covariate set and model specification than $\hat{\beta}_z$.\cite{daniel2020making} Therefore, the treatment coefficient $\hat{\beta}_z$ does not target the estimand of interest. 

% An indirect comparison of conditional treatment effects cannot be performed because we do not have access to the $BC$ study patient-level data and cannot derive a compatible conditional estimate for $B$ vs.~$C$, using the same outcome regression specification. In any case, a comparison of conditional treatment effects is not of interest when making decisions at the population level in HTA.\cite{remiro2020conflating} 

With non-collapsible effect measures such as the odds ratio in logistic regression, marginal and conditional estimands for non-null effects do not coincide,\cite{janes2010quantifying} even with covariate balance and in the absence of confounding.\cite{greenland1987interpretation, greenland1999confounding} Targeting the wrong estimand may induce systematic bias, as observed in a recent simulation study.\cite{remiro2020methods}  Most applications of population-adjusted indirect comparisons are in oncology\cite{remiro2020methods,phillippo2019population} and are concerned with non-collapsible measures of treatment effect such as (log) hazard ratios\cite{austin2011introduction,greenland1987interpretation, austin2014use} or (log) odds ratios.\cite{austin2011introduction, greenland1987interpretation,greenland1999confounding, austin2014use} With both collapsible and non-collapsible measures of effect, maximum-likelihood estimators targeting distinct estimands will have different standard errors. Therefore, marginal and conditional estimates quantify parametric uncertainty differently, and conflating these will lead to the incorrect propagation of uncertainty to the wider health economic decision model, which will be problematic for probabilistic sensitivity analyses. 

\subsection{Marginalization via parametric G-computation}\label{subsec34}

The crucial element that has been missing from the typical application of outcome regression is the marginalization of the $A$ vs. $C$ treatment effect estimate. When adjusting for covariates, one must integrate or average the conditional estimate over the joint $BC$ covariate distribution to recover a marginal treatment effect that is compatible in the indirect comparison. Parametric G-computation\cite{robins1986new, robins1987graphical, keil2018bayesian} is an established method for marginalizing regression-adjusted conditional estimates. We discuss how this methodology can be used in population-adjusted indirect comparisons. 

G-computation in this context consists of: (1) predicting the conditional outcome expectations under treatments $A$ and $C$ for each subject in the $BC$ population; (2) averaging the predictions to produce marginal outcome means on the natural scale; and (3) back-transforming the averages to the linear predictor scale, contrasting the linear predictions to estimate the marginal $A$ vs.~$C$ treatment effect in the $BC$ population. This marginal effect is compatible in the indirect treatment comparison. This procedure is a form of standardization, a technique which has been performed for decades in epidemiology, e.g.~when computing standardized mortality ratios.\cite{vansteelandt2011invited} Parametric G-computation is often called model-based standardization\cite{moore2009covariate, austin2010absolute} because a parametric model is used to predict the conditional outcome expectations under each treatment. When the covariates and outcome are discrete, the estimation of the conditional expectations could be non-parametric, in which case G-computation is numerically identical to crude direct post-stratification.\cite{miettinen1972standardization} 

G-computation marginalizes the conditional estimates by separating the regression modeling outlined in subsection \ref{subsec33} from the estimation of the marginal treatment effect for $A$ vs.~$C$. Firstly, a regression model of the observed outcome $\boldsymbol{y}$ on the covariates $\boldsymbol{x}$ and treatment $\boldsymbol{z}$ is fitted to the $AC$ IPD:
\begin{equation}
g(\mu_n) = \beta_0 +  \boldsymbol{x}_n\boldsymbol{\beta_1} + \left(\beta_z +  \boldsymbol{x}_n^{\boldsymbol{(EM)}} \boldsymbol{\beta_2}\right)\mathds{1}(z_n=1).
\label{eqn3}
\end{equation}
In the context of G-computation, this regression model is often called the ``Q-model''. Contrary to Equation \ref{eqn2}, it is not centered on the mean $BC$ covariates. 

Having fitted the Q-model, the regression coefficients are treated as nuisance parameters. The parameters are applied to the simulated covariates $\boldsymbol{x^*}$ to predict hypothetical outcomes for each subject under both possible treatments. Namely, a pair of predicted outcomes, also called \textit{potential} outcomes,\cite{imbens2015causal} under $A$ and under $C$, is generated for each subject.

% Because G-computation has been developed within the counterfactual framework for causal inference,\cite{robins1986new} we refer to these outcomes as counterfactual outcomes. In this article, these are known as counterfactual to denote what outcomes might have been observed had subjects in a different population, in which the $A$ vs.~$C$ trial was not conducted, received treatment.

Parametric G-computation typically relies on maximum-likelihood estimation to fit the regression model in Equation \ref{eqn3}. In this case, the methodology proceeds as follows. We denote the maximum-likelihood estimate of the regression parameters as  $\boldsymbol{\hat{\beta}} = (\hat{\beta}_0, \boldsymbol{\hat{\beta}_1}, \boldsymbol{\hat{\beta}_2}, \hat{\beta}_z)$. Leaving the simulated covariates $\boldsymbol{x^*}$ at their set values, we fix the treatment values, indicated by a vector $\boldsymbol{z^*} = (z^*_1, z^*_2, \dots, z^*_{N^*})$, for all $N^*$. By plugging treatment $A$ into the maximum-likelihood regression fit for each simulated individual, we predict the marginal outcome mean, on the natural scale, when all subjects are under treatment $A$:

\begin{align}
\hat{\mu}_1
&= \int_{\boldsymbol{x^*}} g^{-1}(\hat{\beta}_0 +  \boldsymbol{x^*}\boldsymbol{\hat{\beta}_1} + \hat{\beta}_z +  \boldsymbol{x^{*(EM)}}\boldsymbol{\hat{\beta}_2}) p(\boldsymbol{x^*}) d\boldsymbol{x^*} \label{eqn4}\\
&\approx
\frac{1}{N^*}\sum_{i=1}^{N^*} g^{-1}  (\hat{\beta}_0 +  \boldsymbol{x}_i^*\boldsymbol{\hat{\beta}_1} + \hat{\beta}_z +  \boldsymbol{x}_i^{*\boldsymbol{(EM)}} \boldsymbol{\hat{\beta}_2}).
\label{eqn5}
\end{align}
Equation \ref{eqn4} follows directly from the law of total expectation. The joint probability density function for the $BC$ covariates is denoted $p(\boldsymbol{x}^*)$, and could be replaced by a probability mass function if the covariates are discrete, or by a mixture density if there is a combination of discrete and continuous covariates. Replacing the integral by the summation in Equation \ref{eqn5} follows from using the empirical joint distribution of the simulated covariates as a non-parametric estimator of the density $p(\boldsymbol{x}^*)$.\cite{daniel2020making} 

Similar to above, by plugging treatment $C$ into the regression fit for every simulated observation, we predict the marginal outcome mean in the hypothetical scenario in which all units are under treatment $C$:

\begin{align}
\hat{\mu}_0
&= \int_{\boldsymbol{x^*}} g^{-1}(\hat{\beta}_0 +  \boldsymbol{x^*}\boldsymbol{\hat{\beta}_1}) p(\boldsymbol{x^*}) d\boldsymbol{x^*} \label{eqn6} \\
&\approx \frac{1}{N^*}\sum_{i=1}^{N^*} g^{-1}(\hat{\beta}_0 +  \boldsymbol{x}_i^*\boldsymbol{\hat{\beta}_1}). 
\label{eqn7} 
\end{align}
To estimate the marginal or population-average treatment effect for $A$ vs.~$C$ in the linear predictor scale, one back-transforms to this scale the average predictions, taken over all subjects on the natural outcome scale, and calculates the difference between the average linear predictions:

\begin{equation}
\hat{\Delta}_{10}^{(2)} = g(\hat{\mu}_1) - g(\hat{\mu}_0).
\label{eqn8}
\end{equation}
If the outcome model in Equation \ref{eqn3} is correctly specified, the estimators of the marginal outcome means under each treatment should be consistent with respect to convergence to their true value, and so should the marginal treatment effect estimate. 

% This is provided that the $BC$ population is correctly specified and other assumptions for valid population adjustment in $BC$ are met (see Appendix B of the Supplementary Material). 

For illustrative purposes, consider a logistic regression for binary outcomes. In this case, $\hat{\mu}_1$ is the average of the individual probabilities predicted by the regression when all participants are assigned to treatment $A$. Similarly, $\hat{\mu}_0$ is the average probability when everyone is assigned to treatment $C$. The inverse link function $g^{-1}(\cdot)$ would be the inverse logit function $\expit(\cdot)=\exp(\cdot)/[1+\exp(\cdot)]$, and the average predictions in the probability scale could be substituted into Equation \ref{eqn8} and transformed to the log-odds ratio scale, using the logit link function. More interpretable summary measures of the marginal contrast, e.g.~odds ratios, relative risks or risk differences, can also be produced by manipulating the average natural outcome means differently than in Equation \ref{eqn8}, mapping these to other scales. For instance, a marginal odds ratio can be estimated as $\exp[g(\hat{\mu}_1)]/\exp[g(\hat{\mu}_0)]=\frac{\hat{\mu}_1/(1-\hat{\mu}_1)}{\hat{\mu}_0/(1-\hat{\mu}_0)}$, where $g(\cdot)$ denotes the logit link function. The standard scale commonly used for performing indirect treatment comparisons is the log-odds ratio scale\cite{dias2013evidence, bucher1997results, phillippo2018methods} and this linear predictor scale is used to define effect modification, which is scale-specific.\cite{phillippo2016nice} Hence, we assume that the marginal log-odds ratio is the relative effect measure of interest. 

Note that the estimated absolute outcomes $\hat{\mu}_1$ and $\hat{\mu}_0$, e.g.~the average outcome probabilities under each treatment in the case of logistic regression, are sometimes desirable in health economic models without any further processing.\cite{phillippo2021target} In addition, these could be useful in unanchored comparisons, where there is no common comparator group included in the analysis, e.g.~if the competitor trial is an RCT without a common control or a single-arm trial evaluating the effect of treatment $B$ alone. In the unanchored case, absolute outcome means are compared across studies as opposed to relative effects. However, unanchored comparisons make very strong assumptions which are largely considered impossible to meet (absolute effects are conditionally constant as opposed to relative effects being conditionally constant).\cite{phillippo2018methods, phillippo2016nice}

% We have postulated a single outcome model for all subjects in the $AC$ IPD, which includes the necessary treatment-covariate interaction terms to capture effect modification over the covariates. Nevertheless, another possible strategy is to fit two outcome models separately for each treatment group in the randomized trial, i.e., to fit one regression to the patients under treatment $A$ and then another regression among the patients under $C$, then predicting the conditional outcome expectations on the entire simulated pseudo-population. In this approach, the model-fitting is performed independently of reference to a conditional treatment effect (the fitted regressions do not have a treatment coefficient), and the estimation of treatment-by-covariate interactions is obviated.\cite{lunceford2004stratification} Throughout the article, we consider the nuisance model in Equation \ref{eqn3} to be a parametric regression. Alternatively, non-parametric estimators of the conditional expectation may be less susceptible to model misspecification.

% \subsubsection{Model fitting and selection}\label{subsec341}

% Because the regression in Equation \ref{eqn3} will be our working model from now onward, we briefly discuss some good practices for model fitting and model selection. Time and care should be taken to perform these exercises and fit an appropriate regression.

The inclusion of all imbalanced effect modifiers in Equation \ref{eqn3} is required for unbiased estimation of both the marginal and conditional $A$ vs.~$C$ treatment effects in the $BC$ population.\cite{zhang2016new} A strong fit of the regression model, evaluated by model checking criteria such as the residual deviance and information criteria, may increase precision. Hence, we could select the model with the lowest information criterion conditional on including all effect modifiers.\cite{zhang2016new} Model checking criteria should not guide causal decisions on effect modifier status, which should be defined prior to fitting the outcome model. As effect-modifying covariates are likely to be good predictors of outcome, the inclusion of appropriate effect modifiers should provide an acceptable fit. In addition, note that any model comparison criteria will only provide information about the observed $AC$ data and therefore tell just part of the story. We have no information on the fit of the selected model to the $BC$ patient-level data. 

In Appendix C of the Supplementary Material, we develop a parametric G-computation approach where the nuisance model is a Cox proportional hazards regression. In this setting, $\hat{\Delta}_{10}^{(2)}$ and $\hat{\Delta}_{20}^{(2)}$ should target marginal log hazard ratios for indirect treatment comparisons in the linear predictor scale. This development is important and useful to practitioners because the most popular outcome types in applications of population-adjusted indirect comparisons are survival or time-to-event outcomes (e.g.~overall or progression-free survival), and the most prevalent measure of effect is the (log) hazard ratio.\cite{phillippo2019population}

From a frequentist perspective, it is not easy to derive analytically a closed-form expression for the standard error of the marginal $A$ vs.~$C$ treatment effect with non-linear outcome models. Deriving the asymptotic distribution is not straightforward as the estimate is a non-linear function of each of the components of $\boldsymbol{\hat{\beta}}$. When using maximum-likelihood estimation to fit the outcome model, standard errors and interval estimates can be obtained using resampling-based methods such as the traditional non-parametric bootstrap\cite{efron1986bootstrap} or the m-out-of-n bootstrap.\cite{varadhan2016cross} In our bootstrap implementation, we only resample the IPD of the $AC$ trial due to patient-level data limitations for the $BC$ study. The standard error would be estimated as the sample standard deviation of the resampled marginal treatment effect estimates. Assuming that the sample size $N$ is reasonably large, we can appeal to the asymptotic normality of the marginal treatment effect and construct Wald-type normal distribution-based confidence intervals. Alternatively, one can construct interval estimates using the relevant quantiles of the bootstrapped treatment effect estimates, without necessarily assuming normality. This avoids relying on the adequacy of the asymptotic normal approximation, an approximation which will be inappropriate where the true model likelihood is distinctly non-normal,\cite{rubin1987logit} and may allow for the more principled propagation of uncertainty.

An alternative to bootstrapping for statistical inference is to simulate the parameters of the multivariable regression in Equation \ref{eqn3} from the asymptotic multivariate normal distribution with means set to the maximum-likelihood estimator and with the corresponding variance-covariance matrix, iterate over Equations \ref{eqn4}-\ref{eqn8} and compute the sample variance. This parametric simulation approach is less computationally intensive than bootstrap resampling. It has the same reliance on random numbers and may offer similar performance.\cite{aalen1997markov} It is equivalent to approximating the posterior distribution of the regression parameters, assuming constant non-informative priors and a large enough sample size. Again, this large-sample formulation relies on the adequacy of the asymptotic normal approximation. 

The choice of the number of bootstrap resamples is important. Given recent advances in computing power, we encourage setting this value as large as possible, in order to maximize the precision and accuracy of the treatment effect estimator, and to minimize the Monte Carlo error in the estimate. A sensible strategy is to increase the number of bootstrap resamples until repeated analyses across different random seeds give similar results, within a specified degree of accuracy.

\subsection{Bayesian parametric G-computation}\label{subsec35}

A Bayesian approach to parametric G-computation may be beneficial for several reasons. Firstly, the maximum-likelihood estimates of the outcome regression coefficients may be unstable where the sample size $N$ of the $AC$ IPD is small, the data are sparse or the covariates are highly correlated, e.g.~due to finite-sample bias or variance inflation. This leads to poor frequentist properties in terms of precision. A Bayesian approach with default shrinkage priors, i.e., priors specifying a low likelihood of a very large effect, can reduce variance, stabilize the estimates and improve their accuracy in these cases.\cite{keil2018bayesian} 

Secondly, we can use external data and/or contextual information on the prognostic effect and effect-modifying strength of covariates, e.g.~from covariate model parameters reported in the literature, to construct informative prior distributions for $\boldsymbol{\beta_1}$ and $\boldsymbol{\beta_2}$, respectively, and skeptical priors (i.e., priors with mean zero, where the variance is chosen so that the probability of a large effect is relatively low) for the conditional treatment effect $\beta_z$, if necessary. Where meaningful prior knowledge cannot be leveraged, one can specify generic default priors instead. For instance, it is unlikely in practice that conditional odds ratios are outside the range $0.1-10$. Therefore, we could use a null-centered normal prior with standard deviation 1.15, which is equivalent to just over 95\% of the prior mass being between 0.1 and 10. As mentioned earlier, this ``weakly informative'' contextual knowledge may result in shrinkage that improves accuracy with respect to maximum-likelihood estimators.\cite{keil2018bayesian}  Finally, it is simpler to account naturally for issues in the $AC$ IPD such as missing data and measurement error within a Bayesian formulation.\cite{keil2014autism, josefsson2021bayesian}

In the generalized linear modeling context, consider that we use Bayesian methods to fit the outcome regression model in Equation \ref{eqn3}. The difference between Bayesian G-computation and its maximum-likelihood counterpart is in the estimated distribution of the predicted outcomes. The Bayesian approach also marginalizes, integrates or standardizes over the joint posterior distribution of the conditional nuisance parameters of the outcome regression, as well as the joint covariate distribution $p(\boldsymbol{x^*})$. Following Keil et al.,\cite{keil2018bayesian} Rubin\cite{rubin1978bayesian} and Saarela et al.,\cite{saarela2015predictive} we draw a vector of size $N^*$ of predicted outcomes $\boldsymbol{y}^*_{z^*}$ under each set intervention $z^* \in \{0,1\}$ from its posterior predictive distribution under the specific treatment. This is defined as $p(\boldsymbol{y}^*_{z^*} \mid \mathcal{D}_{AC})=\int_{\boldsymbol{\beta}} p(\boldsymbol{y}^*_{z^*} \mid \boldsymbol{\beta}) p(\boldsymbol{\beta} \mid \mathcal{D}_{AC}) d\boldsymbol{\beta}$, where $p(\boldsymbol{\beta} \mid \mathcal{D}_{AC})$ is the posterior distribution of the outcome regression coefficients $\boldsymbol{\beta}$, which encode the predictor-outcome relationships observed in the $AC$ trial IPD. This\cite{keil2018bayesian} is given by:
\begin{align}
p(\boldsymbol{y}^*_{z^*} \mid \mathcal{D}_{AC}) &= \int_{\boldsymbol{x^*}} p(\boldsymbol{y}^* \mid \boldsymbol{z^*}, \boldsymbol{x^*}, \mathcal{D}_{AC}) p(\boldsymbol{x^*} \mid \mathcal{D}_{AC}) d\boldsymbol{x^*} 
\label{eqn14}
\\
&= \int_{\boldsymbol{x^*}} \int_{\boldsymbol{\beta}} p(\boldsymbol{y^*} \mid \boldsymbol{z^*}, \boldsymbol{x^*},\boldsymbol{\beta}) p(\boldsymbol{x^*} \mid \boldsymbol{\beta}) p (\boldsymbol{\beta} \mid \mathcal{D}_{AC}) d \boldsymbol{\beta}  d \boldsymbol{x^*}.
\label{eqn15}
\end{align}
As noted by Keil et al.,\cite{keil2018bayesian} the posterior predictive distribution $p(\boldsymbol{y}^*_{z^*} \mid \mathcal{D}_{AC})$ is a function only of the observed data $\mathcal{D}_{AC}$, the joint probability density function $p(\boldsymbol{x^*})$ of the simulated $BC$ pseudo-population, which is independent of $\boldsymbol{\beta}$, the set treatment values $\boldsymbol{z^*}$, and the prior distribution $p(\boldsymbol{\beta})$ of the regression coefficients.

In practice, the integrals in Equations \ref{eqn14} and \ref{eqn15} can be approximated numerically, using full Bayesian estimation via Markov chain Monte Carlo (MCMC) sampling. This is carried out as follows. As per the maximum-likelihood procedure, we leave the simulated covariates at their set values and fix the value of treatment to create two datasets: one where all simulated subjects are under treatment $A$ and another where all simulated subjects are under treatment $C$. The outcome regression model in Equation \ref{eqn3} is fitted to the original $AC$ IPD with the treatment actually received. From this model, conditional parameter estimates are drawn from their posterior distribution $p(\boldsymbol{\beta} \mid \mathcal{D}_{AC})$, given the observed patient-level data and some suitably defined prior $p(\boldsymbol{\beta})$. 

It is relatively straightforward to integrate the model-fitting and outcome prediction within a single Bayesian computation module using efficient simulation-based sampling methods such as MCMC. Assuming convergence of the MCMC algorithm, we form realizations of the parameters $ \{ \hat{\boldsymbol{\beta}}^{(l)} = (\hat{\beta}_0^{(l)}, \boldsymbol{\hat{\beta}}^{(l)}_{\boldsymbol{1}},
\boldsymbol{\hat{\beta}}^{(l)}_{\boldsymbol{2}}, \hat{\beta}^{(l)}_z,): l=1,2,\dots, L \}$, where $L$ is the number of MCMC draws after convergence and $l$ indexes each specific draw. Again, these conditional coefficients are nuisance parameters, not of direct interest in our scenario. Nevertheless, the samples are used to extract draws of the conditional expectations for each simulated subject $i$ (the posterior draws of the linear predictor transformed by the inverse link function) from their posterior distribution. The $l$-th draw of the conditional expectation for simulated subject $i$ set to treatment $A$ is:
\begin{equation}
\hat{\mu}_{1,i}^{(l)}=g^{-1}(\hat{\beta}_0^{(l)} +  \boldsymbol{x}_i^*\boldsymbol{\hat{\beta}}^{(l)}_{\boldsymbol{1}} + \hat{\beta}^{(l)}_z +  \boldsymbol{x}_i^{*\boldsymbol{(EM)}} \boldsymbol{\hat{\beta}}^{(l)}_{\boldsymbol{2}}).
\label{eqn16}
\end{equation}
Similarly, the $l$-th draw of the conditional expectation for simulated subject $i$ under treatment $C$ is:
\begin{equation}
\hat{\mu}_{0,i}^{(l)}=g^{-1}(\hat{\beta}_0^{(l)} +  \boldsymbol{x}_i^*\boldsymbol{\hat{\beta}}^{(l)}_{\boldsymbol{1}}). 
\label{eqn17}
\end{equation}

The conditional expectations drawn from Equations \ref{eqn16} and \ref{eqn17} are used to impute the individual-level outcomes $\{ y_{1,i}^{*(l)} : i=1,\dots,N^*; l=1,2,\dots, L\}$ under treatment $A$ and $\{ y_{0,i}^{*(l)} : i=1,\dots,N^*; l=1,2,\dots, L\}$ under treatment $C$, as independent draws from their posterior predictive distribution at each iteration of the MCMC chain. For instance, if the outcome model is a normal linear regression with a Gaussian likelihood, one multiplies the simulated covariates and the set treatment $z^*_i$ for each subject $i$ by the $l$-th random draw of the posterior distribution of the regression coefficients, given the observed IPD and some suitably defined prior, to form draws of the conditional expectation $\hat{\mu}^{(l)}_{z^*,i}$ (which is equivalent to the linear predictor because the link function is the identity link in linear regression). Then each predicted outcome $y_{z^*,i}^{*(l)}$ would be drawn from a normal distribution with mean equal to $\hat{\mu}^{(l)}_{z^*,i}$ and standard deviation equal to the corresponding posterior draw of the error standard deviation. With a logistic regression as the outcome model, one would impute values of a binary response $y_{z^*,i}^{*(l)}$ by random sampling from a Bernoulli distribution with mean equal to the expected conditional probability $\hat{\mu}^{(l)}_{z^*,i}$.

Producing draws from the posterior predictive distribution of outcomes is fairly simple using dedicated Bayesian software such as \texttt{BUGS},\cite{lunn2012bugs} \texttt{JAGS}\cite{plummer2003jags} or \texttt{Stan},\cite{carpenter2017stan} where the outcome regression and prediction can be implemented simultaneously in the same module. Over the $L$ MCMC draws, these programs typically return a $L \times N^*$ matrix of simulations from the posterior predictive distribution of outcomes. The $l$-th row of this matrix is a vector of outcome predictions of size $N^*$ using the corresponding draw of the regression coefficients from their posterior distribution. We can estimate the marginal treatment effect for $A$ vs.~$C$ in the $BC$ population by: (1) averaging out the imputed outcome predictions in each draw over the simulated subjects, i.e., over the columns, to produce the marginal outcome means on the natural scale; and (2) taking the difference in the sample means under each treatment in a suitably transformed scale. Namely, for the $l$-th draw, the $A$ vs.~$C$ marginal treatment effect is:
\begin{equation}
\hat{\Delta}_{10}^{(2,l)} = g \Bigg (\frac{1}{N^*} \sum_{i=1}^{N^*} y^{*(l)}_{1,i} \Bigg ) - g \Bigg (\frac{1}{N^*} \sum_{i=1}^{N^*} y^{*(l)}_{0,i} \Bigg ).
\label{eqn18}
\end{equation}

The average, variance and interval estimates of the marginal treatment effect can be derived empirically from draws of the posterior density, i.e., by taking the sample mean, variance and the relevant percentiles over the $L$ draws, which approximate the posterior distribution of the marginal treatment effect. The computational expense of the Bayesian approach to G-computation is expected to be similar to that of the maximum-likelihood version, given that the latter typically requires bootstrapping for uncertainty quantification. Computational cost can be reduced by adopting approximate Bayesian inference methods such as integrated nested Laplace approximation (INLA)\cite{rue2009approximate} instead of MCMC sampling to draw from the posterior predictive distribution of outcomes. 

Note that Equation \ref{eqn18} is the Bayesian version of Equation \ref{eqn8}. Other parameters of interest can be obtained, e.g.~the risk difference by using the identity link function in this equation, but these are typically not of direct relevance in our scenario. Again, where the contrast between two different interventions is not of primary interest, the absolute outcome draws from their posterior predictive distribution under each treatment may be relevant. The average, variance and interval estimates of the absolute outcomes can be derived empirically over the $L$ draws. An argument in favor of a Bayesian approach is that, once the simulations have been conducted, one can obtain a full characterization of uncertainty on any scale of interest.  

\subsection{Indirect treatment comparison}\label{subsec36}

The estimated marginal treatment effect for $A$ vs.~$C$ is typically compared with that for $B$ vs.~$C$ to estimate the marginal treatment effect for $A$ vs.~$B$ in the $BC$ population. This is the indirect treatment comparison in the $BC$ population performed in Equation \ref{eqn1}. 

There is some flexibility in this step. Bayesian G-computation and the indirect comparison can be performed in one step under an MCMC approach. In this case, the estimation of $\Delta_{20}^{(2)}$ would be integrated within the estimation or simulation of the posterior of $\Delta_{10}^{(2)}$, under suitable priors, and a posterior distribution for $\Delta_{12}^{(2)}$ would be generated. This would require inputting as data the available aggregate outcomes for each treatment group in the published $BC$ study, or reconstructing subject-level data from these outcomes. For binary outcomes, event counts from the cells of a $2\times2$ contingency table would be required to estimate probabilities of the binary outcome as the incidence proportion for each treatment (dividing the number of subjects with the binary outcome in a treatment group by the total number of subjects in the group), to then estimate a marginal log-odds ratio for $B$ vs.~$C$. For survival outcomes, one can input patient-level data (with outcome times and event indicators for each subject) reconstructed from digitized Kaplan-Meier curves, e.g.~using the algorithm by Guyot et al.\cite{guyot2012enhanced} 

The advantage of this approach is that it directly generates a full posterior distribution for $\Delta_{12}^{(2)}$. Hence, its output is perfectly compatible with a probabilistic cost-effectiveness model. Samples of the posterior are directly incorporated into the decision analysis, so that the relevant economic measures can be evaluated for each sample without further distributional assumptions.\cite{dias2013evidence} If necessary, we can take the expectation over the draws of the posterior density to produce a point estimate $\hat{\Delta}_{12}^{(2)}$ of the marginal $A$ vs.~$B$ treatment effect, in the $BC$ population. Variance and interval estimates are derived empirically from the draws. 

Alternatively, we can perform the G-computation and indirect comparison in two steps. Irrespective of the selected inferential framework, point estimates $\hat{\Delta}_{10}^{(2)}$ and $\hat{\Delta}_{20}^{(2)}$ can be directly substituted in Equation \ref{eqn1}. As the associated variance estimates $\hat{V}(\hat{\Delta}_{10}^{(2)})$ and $\hat{V}(\hat{\Delta}_{20}^{(2)})$ are statistically independent, these are summed to estimate the variance of the $A$ vs.~$B$ treatment effect:
\begin{equation}
\hat{V}(\hat{\Delta}_{12}^{(2)}) = \hat{V}(\hat{\Delta}_{10}^{(2)}) + \hat{V}(\hat{\Delta}_{20}^{(2)}).
\label{eqn38}
\end{equation}
With relatively large sample sizes, interval estimates can be constructed using normal distributions, $\hat{\Delta}_{12}^{(2)} \pm 1.96 \sqrt{\hat{V}(\hat{\Delta}_{12}^{(2)})}$. This two-step strategy is simpler and easier to apply but sub-optimal in terms of integration with probabilistic sensitivity analysis, although one could perform forward Monte Carlo simulation from a normal distribution with mean $\hat{\Delta}_{12}^{(2)}$ and variance $\hat{V}(\hat{\Delta}_{12}^{(2)})$. Ultimately, it is the distribution of $\Delta_{12}^{(2)}$ that is relevant for HTA purposes. 

\section{Simulation study}\label{sec4}

\subsection{Aims}\label{subsec41}

The objectives of the simulation study are to benchmark the performance of the different versions of parametric G-computation, and compare it with that of MAIC and the conventional version of STC across a range of scenarios that may be encountered in practice. We evaluate each estimator on the basis of the following finite-sample frequentist characteristics:\cite{morris2019using} (1) unbiasedness; (2) variance unbiasedness; (3) randomization validity;\footnote{In a sufficiently large number of repetitions, $(100\times (1-\alpha))$\% interval estimates based on normal distributions should contain the true value $(100\times(1-\alpha))$\% of the time, for a nominal significance level $\alpha$.} and (4) precision. The selected performance measures assess these criteria specifically (see subsection \ref{subsec45}). The simulation study is reported following the ADEMP (Aims, Data-generating mechanisms, Estimands, Methods, Performance measures) structure.\cite{morris2019using} All simulations and analyses were performed using \texttt{R} software version 3.6.3.\cite{team2013r} The implementation of the methodologies compared in the simulation study, i.e., the ``M'' in ADEMP, is summarized in Table \ref{tab1}. See Appendix A of the Supplementary Material for a more a detailed description of the methods and their specific settings. Example \texttt{R} code applying the methods to a simulated example is provided in Appendix D of the Supplementary Material.\footnote{The files required to run the simulations are available at \url{http://github.com/remiroazocar/Gcomp_indirect_comparisons_simstudy}.}  

\subsection{Data-generating mechanisms}\label{subsec42}

We consider binary outcomes using the log-odds ratio as the measure of effect. The binary outcome may be response to treatment or the occurrence of an adverse event. For trials $AC$ and $BC$, outcome $y_n$ for subject $n$ is simulated from a Bernoulli distribution with probabilities of success generated from logistic regression, such that:
\begin{equation*}
\logit[p(y_n \mid \boldsymbol{x}_n, z_n)] = \beta_0 + \boldsymbol{x}_n \boldsymbol{\beta_1} + (\beta_z +  \boldsymbol{x}_n^{\boldsymbol{(EM)}} \boldsymbol{\beta_2})\mathds{1}(z_n = 1),
\end{equation*}
using the notation of the $AC$ trial data. Four correlated continuous covariates $\boldsymbol{x}_n$ are generated per subject by simulating from a multivariate normal distribution with pre-specified variable means and covariance matrix.\cite{ripley2009stochastic} Two of the covariates are purely prognostic variables; the other two ($\boldsymbol{x}_n^{\boldsymbol{(EM)}}$) are effect modifiers, modifying the effect of both treatments $A$ and $B$ versus $C$ on the log-odds ratio scale, and prognostic variables. 

The strength of the association between the prognostic variables and the outcome is set to $\beta_{1,k} = -\ln (0.5)$, where $k$ indexes a given covariate. This regression coefficient fixes the conditional odds ratio for the effect of each prognostic variable on the odds of outcome at 2, indicating a strong prognostic effect. The strength of interaction of the effect modifiers is set to $\beta_{2,k} = - \ln (0.67)$, where $k$ indexes a given effect modifier.  This fixes the conditional odds ratio for the interaction effect on the odds of the outcome at approximately 1.5. Both active treatments have the same effect modifiers with respect to the common comparator and identical interaction coefficients for each. Therefore, the shared effect modifier assumption\cite{phillippo2016nice} holds in the simulation study by design. Pairwise Pearson correlation coefficients between the covariates are set to 0.2, indicating a moderate level of positive correlation.

% The $A$ vs.~$B$ marginal treatment effect can be generalized to any given target population, because effect modifiers are guaranteed to cancel out (the marginal effect for $A$ vs.~$B$ is conditionally constant across all populations).

% This parameter has a material impact on the marginal $A$ vs.~$B$ treatment effect. Hence, population adjustment is necessary in order to remove the bias induced by covariate imbalances.

The binary outcome represents the occurrence of an adverse event. Each active intervention has a very strong conditional treatment effect $\beta_z = \ln(0.17)$ at baseline (when the effect modifiers are zero) versus the common comparator. Such relative effect is associated with a ``major'' reduction of serious adverse events in a classification of extent categories by the German national HTA agency.\cite{skipka2016methodological} The covariates may represent comorbidities, which are associated with greater rates of the adverse event and, in the case of the effect modifiers, which interact with treatment to render it less effective. The intercept $\beta_0 = -0.62$ is set to fix the baseline event percentage at 35\% (under treatment $C$, when the values of the covariates are zero). 

The number of subjects in the $BC$ trial is 600, under a 2:1 active treatment vs.~control allocation ratio. For the $BC$ trial, the individual-level covariates and outcomes are aggregated to obtain summaries. The continuous covariates are summarized as means and standard deviations, which would be available to the analyst in the published study in a table of baseline characteristics in the RCT publication. The binary outcomes are summarized as overall event counts, e.g.~from the cells of a $2\times2$ contingency table. Typically, the published study only provides this aggregate information to the analyst. 

The simulation study investigates two factors in an arrangement with nine scenarios, thus exploring the interaction between these factors. The simulation scenarios are defined by the values of the following parameters:

\begin{itemize}
\item The number of subjects in the $AC$ trial, $N \in \{200, 400, 600\}$ under a 2:1 active intervention vs.~control allocation ratio. The sample sizes correspond to typical values for a Phase III RCT\cite{stanley2007design} and for trials included in applications of MAIC submitted to HTA authorities.\cite{phillippo2019population}
\item The degree of covariate imbalance.\footnote{In the simulation study, covariate \textit{balance} is a proxy for covariate \textit{overlap}. Imbalance refers to the difference in covariate distributions across studies, as measured by the difference in (standardized) average values. Imbalances in effect measure modifiers across studies induce bias in the standard indirect comparison and motivate the use of population adjustment. Overlap describes how similar the covariate ranges are  across studies --- there is complete overlap if the ranges are identical. Lack of complete overlap hinders the use of population adjustment.} For both trials, each covariate $k$ follows a normal marginal distribution with mean $\mu_k$ and standard deviation $\sigma_k$, such that $x_{i,k} \sim \textnormal{Normal}(\mu_k, \sigma_k^2)$ for subject $i$. For the $BC$ trial, we fix $\mu_k=0.6$. For the $AC$ trial, we vary the means of the marginal normal distributions such that $\mu_k \in \{ 0.45, 0.3, 0.15\}$. The standard deviation of each marginal distribution is fixed at $\sigma_k=0.4$ for both trials. This setup corresponds to standardized differences or Cohen effect size indices\cite{cohen2013statistical} (the difference in means in units of the pooled standard deviation) of 0.375, 0.75 and 1.125, respectively. This yields strong, moderate and poor covariate overlap; with overlap between the univariate marginal distributions of 85\%, 71\% and 57\%, respectively, when $N=600$. To compute the overlap percentages, we have followed a derivation by Cohen\cite{cohen2013statistical} for normally-distributed populations with equal size and equal variance. Note that the percentage overlap between the joint covariate distributions of each study is substantially lower. The strong, moderate and poor covariate overlap scenarios correspond to average percentage reductions in effective sample size of 22\%, 60\% and 85\%, respectively. These percentage reductions are representative of the range encountered in NICE technology appraisals.\cite{remiro2020methods, phillippo2019population}
\end{itemize}

\subsection{Estimands}\label{subsec43}

The estimand of interest is the marginal log-odds ratio for $A$ vs.~$B$ in the $BC$ population. The treatment coefficient $\beta_z = \ln(0.17)$ is the same for both $A$ vs.~$C$ and $B$ vs.~$C$, and the shared effect modifier assumption holds in the simulation study. Therefore, the true conditional treatment effect for $A$ vs.~$B$ in the $BC$ population is zero. As the true subject-level conditional effects are zero for all units, the true marginal log-odds ratio in the $BC$ population is zero ($\Delta_{12}^{(2)}=0$). This implies a null hypothesis-like simulation setup of no treatment effect for $A$ vs.~$B$, and marginal and conditional estimands in the $BC$ population coincide by design.

Note that the true marginal effect for $A$ vs.~$B$ in the $BC$ population is a composite of that for $A$ vs.~$C$ and that for $B$ vs.~$C$, both of which are non-null. These are the same and cancel out. For reference, the true marginal log-odds ratio in the $BC$ population for the active treatments vs.~the common comparator ($\Delta_{10}^{(2)}$ and $\Delta_{20}^{(2)}$) is computed as -1.15.\footnote{This has been calculated by simulating two potential cohorts of 500,000 subjects, with the $BC$ covariate distribution and the outcome-generating mechanism in subsection \ref{subsec42}. One cohort is under the active treatment and the other is under the common comparator. The number of simulated subjects is sufficiently large to minimize sampling variability. The two cohorts are concatenated and a simple logistic regression is fitted, regressing the simulated binary outcomes on an indicator variable for treatment assignment. The treatment coefficient estimates the average difference in the potential outcomes on the log-odds ratio scale, and serves as the log of the true marginal odds ratio for the two interventions under consideration. This is because the outcomes have been generated according to the true data-generating mechanism, where the true conditional effects are explicit, and which uses the correct conditional model by definition. Due to the non-collapsibility of the odds ratio, this simulation-based approach is necessary to determine the true marginal effect for $A$ vs. $C$ and $B$ vs. $C$.} All methods perform the same unadjusted analysis (i.e., a simple regression of outcome on treatment) to estimate the marginal treatment effect of $B$ versus $C$. Because the $BC$ study is a relatively large RCT, this comparison should be unbiased with respect to the true marginal log-odds ratio in $BC$. Therefore, any bias in the $A$ vs. $B$ comparison should arise from bias in the $A$ vs. $C$ comparison, for which marginal and conditional relative treatment effects are non-null.

% The $A$ vs.~$B$ marginal treatment effect can be generalized to any given target population, because effect modifiers are guaranteed to cancel out (the marginal effect for $A$ vs.~$B$ is conditionally constant across all populations).

% Both active treatments have identical effect modifiers and interaction coefficients for each effect modifier. Therefore, the shared effect modifier assumption\cite{phillippo2016nice} holds in the simulation study by design. The $A$ vs.~$B$ marginal treatment effect can be generalized to any given target population, because effect modifiers are guaranteed to cancel out (the marginal effect for $A$ vs.~$B$ is conditionally constant across all populations).

\subsection{Performance measures}\label{subsec45}

We generate and analyze 2,000 Monte Carlo replicates of trial data per simulation scenario. In our implementations of MAIC and G-computation, a large number of bootstrap resamples or MCMC draws are performed for each of the 2,000 replicates (see Appendix A of the Supplementary Material). For instance, the analysis for one simulation scenario using Bayesian G-computation contains 4,000 MCMC draws (after burn-in) times 2,000 simulation replicates, which equals a total of 8 million posterior draws. Based on the method and simulation scenario with the highest long-run variability (MAIC with $N=200$ and poor covariate overlap), we consider the degree of precision provided by the Monte Carlo standard errors under 2,000 replicates to be acceptable in relation to the size of the effects.\footnote{Conservatively, we assume that $\textnormal{SD}(\hat{\Delta}_{12}^{(2)}) \leq 1.71$ and that the variance across simulations of the estimated treatment effect is always less than 2.92. Given that the MCSE of the bias is equal to $\sqrt{\textnormal{Var}(\hat{\Delta}_{12}^{(2)})/N_{sim}}$, where $N_{sim}=2000$ is the number of simulations, it is at most 0.038 under 2,000 simulations. We consider the degree of precision provided by the MCSE of the bias to be acceptable in relation to the size of the effects. If the empirical coverage rate of the methods is 95\%, $N_{sim}=2000$ implies that the MCSE of the coverage is $\left (\sqrt{(95 \times 5)/2000} \right)\%=0.49\%$, with the worst-case MCSE being $1.12\%$ under 50\% coverage. We also consider this degree of precision to be acceptable. Hence, the simulation study is conducted under $N_{sim}=2000$.} 

We evaluate the performance of the outcome regression methods and MAIC on the basis of the following criteria: (1) bias; (2) variability ratio; (3) empirical coverage rate of the interval estimates; (4) empirical standard error (ESE); and (5) mean square error (MSE). These criteria are explicitly defined in a previous simulation study by the authors.\cite{remiro2020methods} 

With respect to the simulation study aims in subsection \ref{subsec41}, the bias in the estimated treatment effect assesses aim 1. This is equivalent to the average estimated treatment effect across simulations because the true treatment effect $\Delta_{12}^{(2)} = 0$. The variability ratio evaluates aim 2. This represents the ratio of the average model standard error and the sample standard deviation of the treatment effect estimates (the empirical standard error). Variability ratios greater than (or lesser than) one indicate that model standard errors overestimate (or underestimate) the variability of the treatment effect estimate. It is worth noting that this metric assumes that the correct estimand and corresponding variance are being targeted. A variability ratio of one is of little use if this is not the case, e.g.~if both the model standard errors and the empirical standard errors are taken over estimates targeting the wrong estimand. Coverage targets aim 3, and is estimated as the proportion of simulated datasets for which the true treatment effect is contained within the nominal $(100\times(1 - \alpha))\%$ interval estimate of the estimated treatment effect. In this article, $\alpha=0.05$ is the nominal significance level. The empirical standard error is the standard deviation of the treatment effect estimates across the 2,000 simulated datasets. Therefore, it measures precision or long-run variability, and evaluates aim 4. The mean square error is equivalent to the average of the squared bias plus the variance across the 2,000 simulated datasets. Therefore, it is a summary value of overall accuracy (efficiency), that accounts for both bias (aim 1) and variability (aim 4). 

\section{Results}\label{sec5}

\begin{figure}[!htb]
\center{\includegraphics[width=0.63\textwidth]{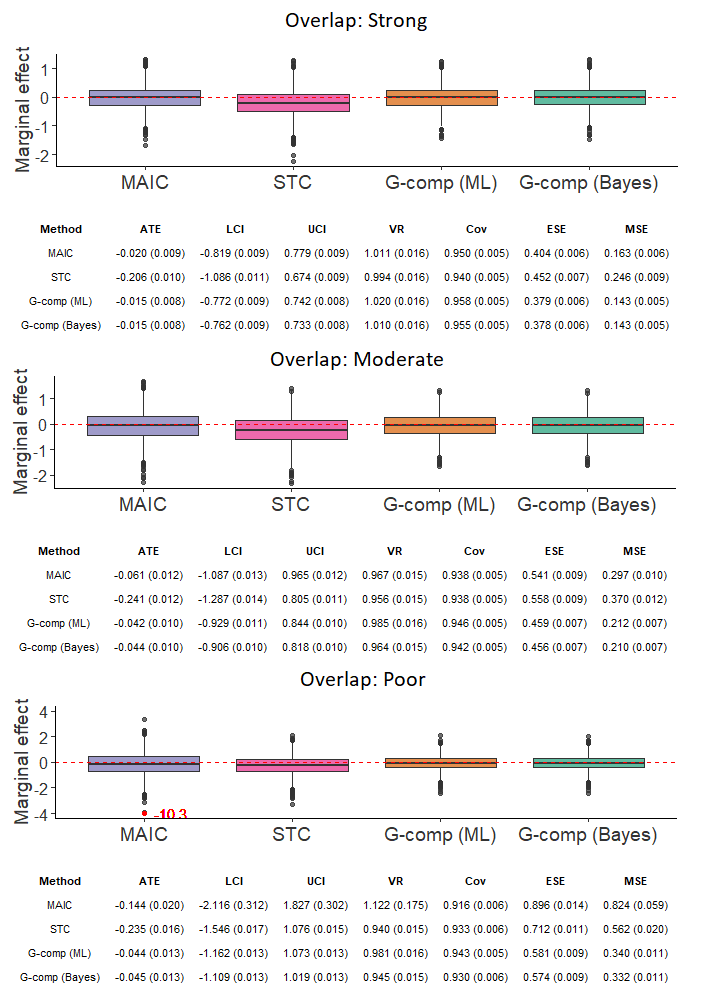}}
\caption{Point estimates and performance metrics across all methods for each simulation scenario with $N=200$. The model standard error for the MAIC outlier in the poor overlap scenario has an inordinate influence on the variability ratio; removing it reduces the variability ratio to 0.980 (0.019). Note that the version of STC evaluated does not actually target a marginal effect. 
\label{fig3}}
\end{figure}

\begin{figure}[!htb]
\center{\includegraphics[width=0.59\textwidth]{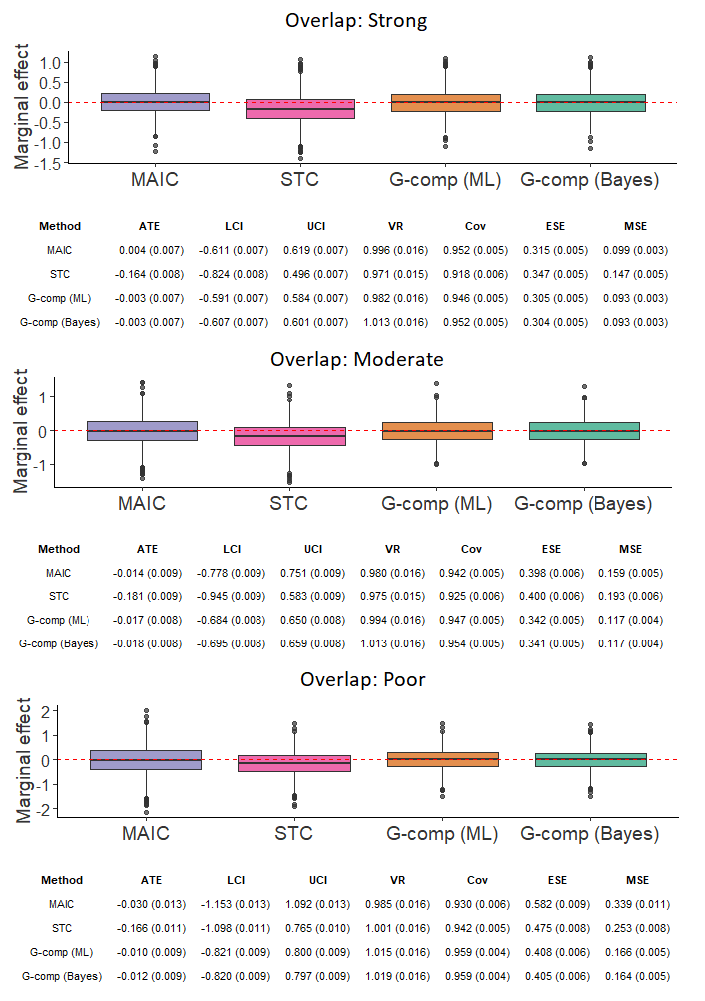}}
\caption{Point estimates and performance metrics across all methods for each simulation scenario with $N=400$.\label{fig4}}
\end{figure}

\begin{figure}[!htb]
\center{\includegraphics[width=0.59\textwidth]{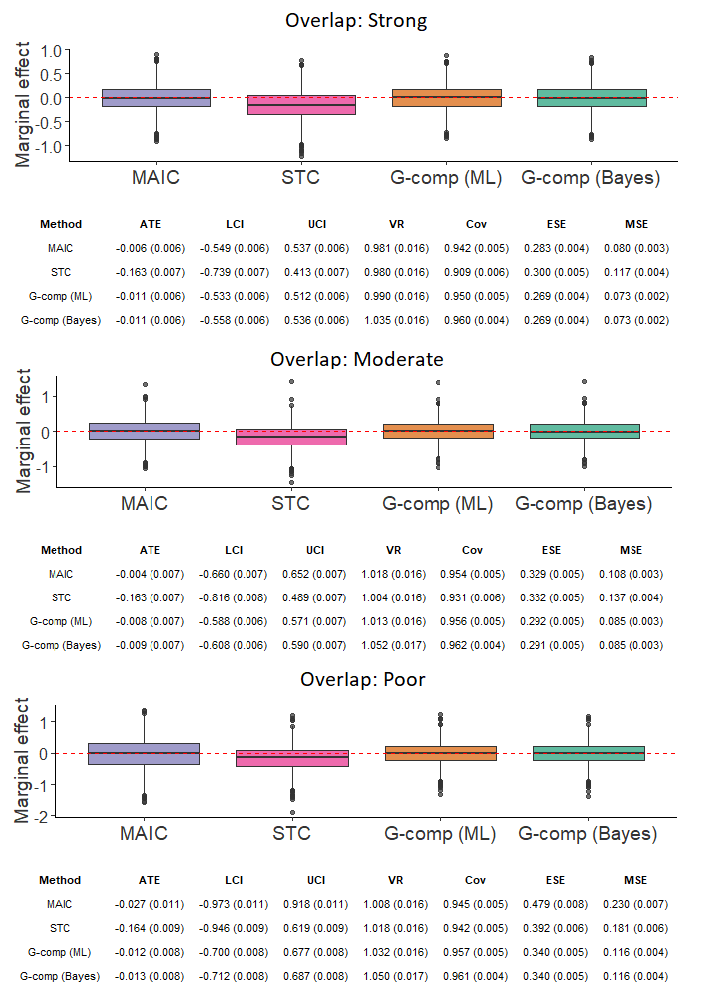}}
\caption{Point estimates and performance metrics across all methods for each simulation scenario with $N=600$.\label{fig5}}
\end{figure}

Performance metrics for all simulation scenarios are displayed in Figure \ref{fig3}, Figure \ref{fig4} and Figure \ref{fig5}. Figure \ref{fig3} displays the results for the three data-generating mechanisms under $N=200$. Figure \ref{fig4} presents the results for the three scenarios with $N=400$. Figure \ref{fig5} depicts the results for the three scenarios with $N=600$. From top to bottom, each figure considers the scenario with strong overlap first, followed by the moderate and poor overlap scenarios. For each scenario, there is a box plot of the point estimates of the $A$ vs.~$B$ marginal treatment effect across the 2,000 simulated datasets. Below, is a summary tabulation of the performance measures for each method. Each performance measure is followed by its Monte Carlo standard error, presented in parentheses, which quantifies the simulation uncertainty. 

In the figures, ATE is the average marginal treatment effect estimate for $A$ vs.~$B$ across the simulated datasets (this is equal to the bias as the true effect is zero). LCI is the average lower bound of the 95\% interval estimate. UCI is the average upper bound of the 95\% interval estimate. VR, ESE and MSE are the variability ratio, empirical standard error and mean square error, respectively. Cov is the empirical coverage rate of the 95\% interval estimates. G-comp (ML) stands for the maximum-likelihood version of parametric G-computation and G-comp (Bayes) denotes its Bayesian counterpart using MCMC estimation. 

Weight estimation cannot be performed for 4 of the 18,000 replicates in MAIC, where there are no feasible weighting solutions. This issue occurs in the most extreme scenario, corresponding to $N=200$ and poor covariate overlap. Feasible weighting solutions do not exist due to separation problems, i.e., there is a total lack of covariate overlap. Because MAIC is incapable of producing an estimate in these cases, the affected replicates are discarded altogether (the scenario in question analyzes 1,996 simulated datasets for MAIC).    

\paragraph{Unbiasedness of treatment effect estimates}

The impact of the bias largely depends on the uncertainty in the estimated treatment effect, quantified by the empirical standard error. We compute standardized biases (bias as a percentage of the empirical standard error). With $N=200$, MAIC has standardized biases of magnitude 11.3\% and 16.1\% under moderate and poor covariate overlap, respectively. Otherwise, the magnitude of the standardized bias is below 10\%. Similarly, under $N=200$, the maximum-likelihood version of parametric G-computation has standardized biases of magnitude 13.3\% and 24.8\% in the scenarios with moderate and poor overlap, respectively. In all other scenarios, the standardized bias has magnitude below 10\%. For Bayesian parametric G-computation, standardized biases never have a magnitude above 10\% and troublesome biases are not produced in any of the simulation scenarios. The maximum absolute value of the standardized bias is 9.7\% in a scenario with $N=200$ and moderate covariate overlap. 

To evaluate whether the bias in MAIC and parametric G-computation has any practical significance, we investigate whether the coverage rates are degraded by it. Coverage is not affected for maximum-likelihood parametric G-computation, where empirical coverage rates for all simulation scenarios are very close to the nominal coverage rate, 0.95 for 95\% interval estimates. In the case of MAIC, there is discernible undercoverage in the scenario with $N=200$ and poor covariate overlap (empirical coverage rate of 0.916). This is the scenario with the lowest effective sample size after weighting. Hence, the results are probably related to small-sample bias in the weighted logistic regression.\cite{vittinghoff2007relaxing} This bias for MAIC was not observed in the more extreme scenarios of a recent simulation study,\cite{remiro2020methods} which considered survival outcomes and the Cox proportional hazards regression as the outcome model. In absolute terms, the bias of MAIC is greater than that of both versions of parametric G-computation where the number of patients in the $AC$ trial is small ($N=200$) or covariate overlap is poor. In fact, when both of these apply, the bias of MAIC is important (-0.144). Otherwise, with $N=400$ or greater, and moderate or strong overlap, the aforementioned methods produce similarly low levels of bias. 

STC generates problematic negative biases in all nine scenarios considered in this simulation study, with a standardized bias of magnitude greater than 30\% in all cases. STC consistently produces the highest bias of all methods, and the magnitude of the bias appears to increase under the smallest sample size ($N=200$). The systematic bias in STC is due to the divergence of the conditional estimates produced for $A$ versus $C$ from the corresponding marginal estimand that should be targeted. This is a result of the non-collapsibility of the (log) odds ratio.

\paragraph{Unbiasedness of variance estimates}

In MAIC, the variability ratio of treatment effect estimates is close to one under all simulation scenarios except one. That is the scenario with $N=200$ and poor covariate overlap, where the variability ratio is 1.122. This high value is attributed to the undue influence of an outlier (as seen in the box plot of point estimates) on the average model standard error. Once the outlier is removed, the variability ratio decreases to 0.98, just outside from being within Monte Carlo error of one but not statistically significantly different. This suggests very little bias in the standard error estimates in this scenario, i.e., that the model standard errors tend to coincide with the empirical standard error. In a previous simulation study,\cite{remiro2020methods} robust sandwich standard errors underestimated the variability of estimates in MAIC under small sample sizes and poor covariate overlap. The non-parametric bootstrap seems to provide more conservative variance estimation in these extreme settings. 

In STC, variability ratios are generally close to one with $N=400$ and $N=600$. Any bias in the estimated variances appears to be negligible, although there is a slight decrease in the variability ratios when the $AC$ sample size is small ($N = 200$). Recall that this metric assumes that the correct estimand and corresponding variance are being targeted. However, in our application of STC, both model standard errors and empirical standard errors are taken over an incompatible indirect treatment comparison.  

In maximum-likelihood parametric G-computation, variability ratios are generally very close to one. In Bayesian parametric G-computation, variability ratios are generally close to one but are slightly above it in some scenarios with $N=600$ (1.05 and 1.052 with moderate and poor covariate overlap, respectively). This suggests some overestimation of the empirical standard error by the model standard errors.

\clearpage

\paragraph{Randomization validity}

From a frequentist viewpoint,\cite{neyman1934two} 95\% interval estimates are randomization-valid if these are guaranteed to include the true treatment effect 95\% of the time. Namely, the empirical coverage rate should be approximately equal to the nominal coverage rate, in this case 0.95 for 95\% interval estimates, to obtain appropriate type I error rates for testing a ``no effect'' null hypothesis. Theoretically, the empirical coverage rate is statistically significantly different to 0.95 if, roughly, it is less than 0.94 or more than 0.96, assuming 2,000 independent simulations per scenario. These values differ by approximately two standard errors from the nominal coverage rate. Poor coverage rates are a decomposition of both the bias and the standard error used to compute the Wald-type interval estimates. In the simulation scenarios, none of the methods lead to overly conservative inferences but there are some issues with undercoverage. 

Empirical coverage rates for MAIC are significantly different from the advertised nominal coverage rate in three scenarios. In the three, the coverage rate is below 0.94 (empirical coverage rates of 0.938, 0.93 and 0.916). The last two of these occur in scenarios with poor covariate overlap, with the latter corresponding to the smallest effective sample size after weighting ($N=200$). This is the scenario which integrates the two most important determinants of small-sample bias, in which MAIC has exhibited discernible bias. In this case, undercoverage is bias-induced. On the other hand, in our previous simulation study,\cite{remiro2020methods} undercoverage was induced by the robust sandwich variance underestimating standard errors.

\clearpage

In the conventional version of STC, coverage rates are degraded by the bias induced by the non-collapsibility of the log-odds ratio. Almost invariably, there is undercoverage. Interestingly, the empirical coverage does not markedly deteriorate --- coverage percentages never fall below 90\%, i.e., never at least double the nominal rate of error. In general, both versions of parametric G-computation exhibit appropriate coverage. Only one scenario provides rates below 0.94 (Bayesian G-computation with $N=200$ and poor overlap, with an empirical coverage rate of 0.93). No scenarios have empirical coverage above 0.96. Coverage rates for the maximum-likelihood implementation are always appropriate, with most empirical coverage percentages within Monte Carlo error of 95\%. 

\paragraph{Precision and efficiency}

Both versions of parametric G-computation have reduced empirical standard errors compared to MAIC across all scenarios. Interestingly, conventional STC is even less precise than MAIC in most scenarios (all the scenarios with moderate or strong overlap, where reductions in effective sample size after weighting are tolerable). Several trends are revealed upon comparison of the ESEs, and upon visual inspection of the spread of the point estimates in the box plots. As expected, the ESE increases for all methods (i.e., estimates are less precise) as the number of subjects in the $AC$ trial is lower. The decrease in precision is more substantial for MAIC than for the outcome regression methods.  

\clearpage

The degree of covariate overlap has an important influence on the ESE and population adjustment methods incur losses of precision when covariate overlap is poor. Again, this loss of precision is more substantial for MAIC than for the outcome regression approaches. Where overlap is poor, there exists a subpopulation in $BC$ that does not overlap with the $AC$ population. Therefore, inferences in this subpopulation rely largely on extrapolation. Outcome regression approaches require greater extrapolation when the covariate overlap is weaker, thereby incurring a loss of precision. 

Where covariate overlap is strong, both versions of parametric G-computation display very similar ESEs than MAIC. As mentioned earlier, conventional STC offers even lower precision than MAIC in these cases. To illustrate this, consider the scenario with $N=200$ and moderate overlap, where MAIC is expected to have a low effective sample size after weighting and perform comparatively worse than outcome regression. Even in this scenario, MAIC appears to be more precise (empirical standard error of 0.541) than conventional STC (empirical standard error of 0.558). As overlap decreases, precision is reduced more markedly for MAIC compared to the outcome regression methods. Under poor overlap, MAIC considerably increases the ESE compared to the conventional STC.

In MAIC, extrapolation is not even possible. Where covariate overlap is poor, the observations in the $AC$ IPD that are not covered by the ranges of the selected covariates in $BC$ are assigned weights that are very close to zero. The relatively small number of individuals in the overlapping region of the covariate space are assigned inflated weights, dominating the reweighted sample. These extreme weights lead to large reductions in effective sample size and affect very negatively the precision of estimates.

Similar to the trends observed for the ESE, the MSE is also very sensitive to the value of $N$ and to the level of covariate overlap. The MSE decreases for all methods as $N$ and the level of overlap increase. The accuracy of MAIC and the G-computation methods is comparable when the $AC$ sample size is high or covariate overlap is strong. As the $AC$ sample size and overlap decrease, the relative accuracy of MAIC with respect to both approaches to G-computation is markedly reduced. Accuracy for the conventional version of STC is comparatively poor and this is driven by bias.

Where covariate overlap is strong or moderate, the parametric G-computation methods have the highest accuracy, followed by MAIC and STC. Where overlap is poor, both versions of parametric G-computation are considerably more accurate than MAIC, with much smaller mean square errors. MAIC also provides less accurate estimates than STC in terms of mean square error. The variability of estimates under MAIC increases considerably in these scenarios. The precision is sufficiently poor to offset the loss of bias with respect to STC. 

\section{Discussion}\label{sec6}

\paragraph{Summary of results}

The parametric G-computation methods and MAIC can yield unbiased estimates of the marginal $A$ vs.~$B$ treatment effect in the $BC$ population. Conventional STC targets a conditional treatment effect for $A$ vs.~$C$ that is incompatible in the indirect comparison. Bias is produced when the measure of effect is non-collapsible. Across all scenarios, both versions of G-computation largely eliminate the bias induced by effect modifier imbalances. There is some negative bias in MAIC and parametric G-computation where the sample size $N$ is small. In the case of MAIC, this is problematic where covariate overlap is poor. MAIC did not display these biases using Cox proportional hazards regression as the outcome model in the time-to-event setting.\cite{remiro2020methods} The difference in results is likely due to logistic regression being more prone to small-sample bias than Cox regression.\cite{vittinghoff2007relaxing} 

As for precision, the G-computation approaches have reduced variability compared to MAIC. The superior precision is demonstrated by their lower empirical standard errors across all scenarios. Because the methods are generally unbiased, precision is the driver of comparative accuracy. The simulation study confirms that, under correct model specification, parametric G-computation methods have lower mean square errors than weighting and are therefore more efficient. The differences in performance are exacerbated where covariate overlap is poor and sample sizes are low. In these cases, the effective sample size after weighting is small, and this leads to inflated variances and wider interval estimates for MAIC.

For the conventional STC, outcome regression may have decreased precision relative to MAIC, as dictated by the empirical standard errors. On the other hand, the marginalized outcome regression methods are more precise than both MAIC and conventional STC. From a frequentist perspective, the standard error of the estimator of a conditional log-odds ratio for $A$ vs.~$C$, targeted by conventional STC, is larger than the standard error of a regression-adjusted estimate of the marginal log-odds ratio for $A$ vs.~$C$, produced by G-computation. This precision comparison likely lacks relevance, because one is comparing estimators that target different estimands. Nevertheless, it supports previous findings on non-collapsible measures of effect when adjusting for prognostic covariates.\cite{daniel2020making, moore2009covariate} When we marginalize and compare estimators targeting like-for-like marginal estimands, we find that outcome regression is no longer detrimental for precision and efficiency compared to weighting.  

The robust sandwich variance estimator for MAIC underestimates variability and produces narrow interval estimates under small effective sample sizes.\cite{remiro2020methods} This simulation study demonstrates that the bootstrap procedure provides more conservative variance estimation in the more extreme settings. This implies that the bootstrap approach should be preferred for statistical inference where there are violations of the overlap assumption and small sample sizes. 

\paragraph{Extrapolation capabilities and precision considerations}

Some care is required to translate the results of this simulation study into real applications, where effective sample sizes and percentage reductions in effective sample size may be lower and higher, respectively, than those considered.\cite{phillippo2019population} In these situations, covariate overlap is poor and this leads to a high loss of precision in MAIC. G-computation methods should be considered because they are substantially more statistically efficient. This is particularly the case where the outcome model is a logistic regression, more prone to small-sample bias,\cite{phillippo2020assessing, vittinghoff2007relaxing} imprecision and inefficiency\cite{annesi1989efficiency} than other models, e.g.~the Cox regression. In addition, where sample sizes are small and the number of covariates is large, feasible weighting solutions may not exist for MAIC due to separation problems.\cite{jackson2020alternative} This is observed in one of the scenarios of this simulation study ($N=200$ with poor overlap) and elsewhere.\cite{phillippo2020assessing} An advantage of outcome regression is that it can be applied in these settings. MAIC cannot extrapolate beyond the covariate space observed in the IPD. Therefore, it cannot overcome the failure of assumptions that is the lack of covariate overlap and is incapable of producing an estimate.

Moreover, we note that MAIC requires accounting for all effect modifiers (balanced and imbalanced), as excluding balanced covariates from the weighting procedure does not ensure balance after the weighting. On the other hand, outcome regression methods do not necessarily require the inclusion of the effect modifiers that are in balance, for instance when the outcome model is a linear regression. This may mitigate losses of precision further, particularly where the number of potential effect modifiers is large. 

With limited overlap, outcome regression methods can use the linearity assumption to extrapolate beyond the $AC$ population, provided the true relationship between the covariates and the outcome is adequately captured. We view this as a desirable attribute because poor overlap, with small effective sample sizes and large percentage reductions in effective sample size, is a pervasive issue in health technology appraisals.\cite{phillippo2019population} Nevertheless, where overlap is more considerable, one may wish to restrict inferences to the region of overlap and avoid relying on a model for extrapolation outside this region.\cite{ho2007matching} 

It is worth noting that we have used the standard MAIC formulation proposed by Signorovitch et al.\cite{signorovitch2010comparative, phillippo2016nice, remiro2020methods,  phillippo2020equivalence} and that our conclusions are based on this approach. Nevertheless, MAIC is a rapidly developing methodology with novel implementations. An alternative formulation based on entropy balancing has been presented.\cite{petto2019alternative,belger2015inclusion, phillippo2020equivalence} This approach is similar to the original version with a subtle modification to the weight estimation procedure. While it has some interesting computational properties, Phillippo et al.\cite{phillippo2020equivalence} have recently shown that the standard method of moments and entropy balancing produce weights that are mathematically equivalent (up to optimization error or a normalizing constant). Jackson et al.\cite{jackson2020alternative} propose a distinct weight estimation procedure that satisfies the conventional method of moments and maximizes the effective sample size. A larger effective sample size translates into minimizing the variance of the weights, with more stable weights producing a gain in precision at the expense of introducing some bias. 

\paragraph{Model specification assumptions}

Note that the model extrapolation uncertainty is not reflected in the interval estimates for the outcome regression approaches and that some consider weighting approaches to give a ``more honest reflection of the overall uncertainty''.\cite{vansteelandt2011invited} The gain in efficiency produced by outcome regression must be counterbalanced against the potential for model misspecification bias. 

This simulation study only considers a best-case scenario with correct parametric model specification. To predict the outcomes in G-computation, we use the logistic regression model implied by the data-generating mechanism. In real applications, parametric modeling assumptions are difficult to hold because, unlike in simulations, the correct specification is unknown, particularly where there are many covariates and complex relationships exist between them. For instance, assumptions will not hold if only linear relationships are considered and the selected covariates have non-linear interactions with treatment on the linear predictor scale. 

Weighting methods are often perceived to rely on less demanding parametric assumptions, yet model misspecification is also a potential issue. Note that, even though the data-generating mechanism of the simulation study does not specify a trial assignment model, the logistic regression for weight estimation is the ``best-case'' model because the ``right'' subset of covariates is selected as effect modifiers.\footnote{The MAIC implementation is optimal in terms of precision and accuracy because the trial assignment model only balances the two covariates that interact with treatment. Nevertheless, these are not the only two covariates that are associated with trial assignment. Consider balancing the full set of covariates that predict trial assignment (a total of four covariates, including the two predictors with only main effects in the data-generating outcome model). Variance would be increased without improving the potential for bias reduction in the $BC$ population. The behavior of MAIC would be more unstable because of weaker overlap. More extreme weights would be produced, and finite-sample or ``chance'' overlap violations would be more likely, particularly with small $AC$ sample sizes.} The balancing property of the weights\cite{dehejia1999causal, waernbaum2010propensity} holds with respect to the effect modifier means --- conditional on the weights, all effect modifier means are balanced between the two studies. More precisely, the estimated weights are scores that, when applied to the $AC$ IPD, form a pseudo-population that has balanced effect modifier means with respect to the $BC$ population. Therefore, there is conditional exchangeability over trial assignment and one can achieve unbiased estimation of treatment effects in the $BC$ population.

The simulation study presented in this article demonstrates proof-of-concept for the outcome regression methods and for MAIC but does not investigate how robust the methods are to failures in assumptions. Future simulation studies should explore performance in scenarios where assumptions are violated, in order to draw more accurate conclusions with respect to practical applications and limitations. Conditional exchangeability (“no omitted effect modifiers”) is a fundamental assumption for all methods. In practice, there may be incomplete information on effect modifiers for one or both trials, and background knowledge or theory about effect modification is often weak. 

The general-purpose nature of the proposed G-computation approaches may provide some degree of robustness against model misspecification because the covariate-adjusted outcome model does not necessarily need to be parametric. Non- and semi-parametric regression techniques or other data-adaptive estimation approaches are viable to detect (higher-order) interactions, product terms and non-linear relationships between regressors, and offer more flexible functions to predict the conditional outcome expectations. These make weaker modeling assumptions than parametric regressions but are more prone to overfitting, particularly with small sample sizes.

\paragraph{Limitations}

Care must be taken where sample sizes are small in population-adjusted indirect comparisons. Low sample sizes cause substantial issues for the accuracy of MAIC due to unstable weights. As the sponsor company is directly responsible for setting the value of $N$, the $AC$ trial should be as large as possible to maximize precision and accuracy. The sample size requirements for indirect comparisons, and more generally for economic evaluation, are considerably larger than those required to demonstrate an effect for the main clinical outcome in a single RCT. However, trials are usually powered for the main clinical comparison, even if there is a prospective indirect, potentially adjusted, comparison down the line. Ideally, if the manufacturer intends to use standard or population-adjusted indirect comparisons for reimbursement purposes, its clinical study should be powered for the relevant methods.

Note that sponsors tend to run multiple RCTs instead of one larger RCT for marketing authorization applications. If there are many different IPD RCTs, it is necessary to fit the covariate-adjusted regression to each patient-level dataset and marginalize against the $BC$ pseudo-population in G-computation. Similarly, one would apply MAIC to each study individually, reweighting each patient-level dataset against the $BC$ study report. Then, a meta-analysis of effect measure estimates can be performed in the same population using the marginalized or weighted results from the IPD studies and the original effect estimate published in the ALD study.

The population adjustment methods outlined in this article are only applicable to pairwise indirect comparisons, and not easily generalizable to larger network structures of treatments and studies. This is because the methods have been developed in the two-study scenario seen in this paper, very common in HTA submissions, where there is one $AC$ study with IPD and another $BC$ study with ALD. In this very sparse network, indirect comparisons are vulnerable to bias induced by effect modifier imbalances. In larger networks, multiple pairwise comparisons do not necessarily generate a consistent set of relative effect estimates for all treatments. This is because the comparisons must be undertaken in the ALD populations.  

Another issue is that the ALD population(s) may not correspond precisely to the target population for the decision. Marginal estimands in different populations may not match if there are differences in the distribution of effect modifiers. This is a problem of external validity: if populations are non-exchangeable, an internally valid estimate of the marginal estimand in one population is not necessarily unbiased for the marginal estimand in the other(s).\cite{westreich2019target, imai2008misunderstandings} To address this, one suggestion would be for the decision-maker to define a target population for a specific disease into which all manufacturers should conduct their indirect comparisons. The outcome regression approaches discussed in this article could be applied to produce marginal effects in any target population. The target could be represented by the joint covariate distribution of a registry, cohort study or some other observational dataset, and one would marginalize over this distribution. Similarly, MAIC can reweight the IPD with respect to a different population than that of the $BC$ study.  

Recently, a novel population adjustment method named multilevel network meta-regression (ML-NMR) has been introduced.\cite{phillippo2020multilevel, phillippo2019calibration} ML-NMR generalizes IPD network meta-regression to include aggregate-level data, reducing to this method when IPD are available for all studies. ML-NMR is a timely addition; it is applicable in treatment networks of any size with the aforementioned two-study scenario as a special case. This is important because a recent review\cite{phillippo2019population} finds that 56\% of NICE technology appraisals include larger networks, where the methods discussed in this article cannot be readily applied. 
 
ML-NMR is an outcome regression approach, with the outcome model of interest being identical to that of parametric G-computation. While the methods share the same assumptions in the two-study scenario, ML-NMR generalizes the regression to handle larger networks. Like Bayesian G-computation, ML-NMR has been developed under a Bayesian framework and estimates the outcome model using MCMC. It also makes parametric assumptions to characterize the marginal covariate distributions in $BC$ and reconstructs the joint covariate distribution using a copula. The methods average over the $BC$ population in different ways; Bayesian G-computation simulates individual-level covariates from their approximate joint distribution and ML-NMR uses numerical integration over the approximate joint distribution (quasi-Monte Carlo methods). 

In its original publication,\cite{phillippo2020multilevel, phillippo2019calibration} ML-NMR targets a conditional treatment effect (avoiding the compatibility issues of conventional STC), because the effect estimate is derived from the treatment coefficient of a covariate-adjusted multivariable regression. However, ML-NMR can directly calculate marginalization integrals akin to those required for Bayesian G-computation (Equations \ref{eqn14} and \ref{eqn15}). Phillippo et al. have recently demonstrated that ML-NMR can be adapted to target marginal treatment effects.\cite{phillippo2021target} We previously mentioned that pairwise population-adjusted indirect comparisons target marginal estimands that are specific to the $BC$ study. These may not be directly relevant for HTA decision-making. On the other hand, ML-NMR can potentially estimate marginal effects in any target population, presenting novel and exciting opportunities for evidence synthesis. 

\paragraph{Concluding remarks}

The traditional regression adjustment approach in population-adjusted indirect comparisons targets a conditional treatment effect for $A$ vs.~$C$. We have showed empirically that this effect is incompatible in the indirect treatment comparison, producing biased estimation where the measure of effect is non-collapsible. In addition, this effect is not of interest in our scenario because we seek marginal effects for policy decisions at the population level. We have proposed approaches for marginalizing the conditional estimates produced by covariate-adjusted regressions. These are applicable to a wide range of outcome models and target marginal treatment effects for $A$ vs.~$C$ that have no compatibility issues in the indirect treatment comparison. 

At present, MAIC is the most commonly used population-adjusted indirect comparison method.\cite{phillippo2019population} We have demonstrated that the novel marginalized outcome regression approaches achieve greater statistical precision than MAIC and are unbiased under no failures of assumptions. Hence, the development of these approaches is appealing and impactful. As observed in the simulation study, outcome regression provides more stable estimators and is more efficient than weighting under correct model specification. This is particularly the case where covariate overlap is poor and effective sample sizes are small. These results support previous findings comparing outcome regression and weighting in different contexts.\cite{vo2021assessing, neugebauer2005prefer, lunceford2004stratification, van2003unified} 

We can now capitalize on the advantages offered by outcome regression with respect to weighting in our scenario, e.g.~extrapolation capabilities and increased precision. Outcome regression methods are also appealing because they make different modeling assumptions than weighting. Sometimes, identifying the variables that affect outcome is more straightforward than identifying the factors that drive trial assignment. This is not typically the case in the standard use of propensity score weighting in observational studies, where one identifies the factors that drive treatment (as opposed to trial) assignment. Nevertheless, in our context, the factors driving selection into different RCTs are often arbitrary. In these cases, the proposed G-computation approaches may be useful. In addition, researchers could develop augmented or doubly robust methods\cite{hernan2020causal, vansteelandt2011invited, bang2005doubly} that combine the outcome model with the trial assignment model for the weights. These approaches are attractive due to their increased flexibility and robustness to model misspecification.\cite{bang2005doubly, kang2007demystifying} 

Furthermore, we have shown that the marginalized regression-adjusted estimates provide greater statistical precision than the conditional estimates produced by the conventional version of STC. While this precision comparison is irrelevant, because it is made for estimators of different estimands, it supports previous research on non-collapsible measures of effect.\cite{daniel2020making, moore2009covariate} 

Marginal and conditional effects are regularly conflated in the literature on population-adjusted indirect comparisons, with many simulation studies comparing the bias, precision and efficiency of estimators of different effect measures. The implications of this conflation must be acknowledged in order to provide meaningful comparisons of methods. We have built on previous research conducted by the original authors of STC, who have also suggested the use of a preliminary ``covariate simulation'' step.\cite{ishak2015simulation,ishak2015simulated} Nevertheless, up until now, there was no consensus on how to marginalize the conditional effect estimates. For instance, in a previous simulation study,\cite{remiro2020methods} we discouraged the ``covariate simulation'' approach when attempting to average on the linear predictor scale. Averaging on the linear predictor scale, i.e., computing the conditional linear prediction under each treatment for every simulated subject and averaging the linear predictions across all subjects, then calculating the difference between the average predictions, reduces to the conventional version of STC (i.e., to ``plugging in'' the mean $BC$ covariate values). As discussed in the next paragraph, it is equivalent to averaging ``predictions at the mean''\cite{bartlett2018covariate} or estimating the ``mean at mean covariates'',\cite{qu2015estimation} hence producing conditional effect estimates for $A$ vs.~$C$, as opposed to marginal estimates. We hope to have established some clarity. 

The readers may notice that the outcome regression fitted in the conventional STC (subsection \ref{subsec33}) is different to that fitted in outcome regression with ``covariate simulation'' (e.g. parametric G-computation in subsection \ref{subsec34}). In the conventional outcome regression described in \ref{subsec33} and by the NICE technical support document,\cite{phillippo2016nice} the IPD covariates are centered by plugging in the mean $BC$ covariate values. In the Q-model required for G-computation, outlined in \ref{subsec34}, the covariates are not centered and the regression fit is used to make predictions for the simulated covariates. The underlying reason for this has been described for generalized linear models with non-linear link functions, such as logistic or Poisson regression.\cite{bartlett2018covariate, qu2015estimation, lane1982analysis} On the natural scale, averaging the individual outcome predictions made at the centered covariates of the sample does not consistently estimate the marginal mean response for the centered sample. In the words of Bartlett,\cite{bartlett2018covariate} ``prediction at the mean'' value of the baseline covariates for a treatment group does not result in the ``marginal mean'' under such treatment. Similarly, in the words of Qu and Luo,\cite{qu2015estimation} the ``mean at mean covariates'' of the study sample is generally not equivalent to the marginal response over the subjects in the sample. The former results in a conditional estimate whereas the latter produces a marginal population-level estimate, of interest in our scenario.

It is important to acknowledge that the methods discussed in this article are required due to a suboptimal scenario, where patient-level data on covariates are unavailable for the $BC$ study. Ideally, these should be made freely available or, at least, made available for the purpose of evidence synthesis by the sponsor company. Raw patient-level data are always the preferred input for statistical inference, allowing for the testing of assumptions.\cite{chan2014increasing} We note that the manufacturer itself could facilitate inference by using the IPD to create fully artificial covariate datasets.\cite{nowok2016synthpop} The release of such datasets would not involve a violation of privacy or confidentiality, avoiding the need for the ``covariate simulation'' step and removing the risk of misspecifying the joint covariate distribution of the $BC$ population. Alternatively, Bonofiglio et al.\cite{bonofigliorecovery} have recently proposed a framework where access to covariate correlation summaries is made possible through distributed computing. It is unclear whether access to such framework would be granted to a competitor submitting evidence for reimbursement to HTA bodies, albeit the summaries could be reported in clinical trial publications.

The presented marginalization methods have been developed in a very specific context, common in HTA, where access to patient-level data is limited and an indirect comparison is required. However, their principles are applicable to estimate marginal treatment effects in any situation. For instance, in scenarios which require marginalizing out regression-adjusted estimates over the study sample in which they have been computed. Alternatively, the frameworks can be used to transport the results of a randomized experiment to any other external target population; not necessarily that of the $BC$ trial. In both cases, the required assumptions are weaker than those required for population-adjusted indirect comparisons. 

Finally, we have assumed that the $AC$ study is a randomized trial. Our approach to parametric G-computation could be extended to the situation where the $AC$ trial is an observational study by including all confounders of the treatment-outcome relationship in the outcome model. In this scenario, one must overcome the limited internal validity of the study design. Because treatment assignment is non-random, additional assumptions would be required, e.g. conditional exchangeability within the study arms (``no unmeasured confounding'') and the associated overlap/positivity condition.\cite{faria2015nice, robins2009estimation} These assumptions are similar to those discussed in subsection \ref{subsec22} but would be expected to hold across treatment arms in the IPD study in addition to across study populations.
 
\section*{Acknowledgments}

All authors contributed to the conception and design of the article, and to the analysis and interpretation of data. All authors drafted the manuscript, critically revised it for important intellectual content, and provided final approval of the version be published. The authors thank the peer reviewers of a previous article of theirs.\cite{remiro2020methods} Their comments were extremely insightful and helped improved the underlying motivation of this article. The authors are hugely thankful to Tim Morris, whose comments on G-computation and marginalization helped motivate the article. Finally, the authors are grateful to David Phillippo, who has provided very valuable feedback to our work, and helped in substantially improving our research. This article is based on research supported by Antonio Remiro-Az\'ocar's PhD scholarship from the Engineering and Physical Sciences Research Council of the United Kingdom. Anna Heath is supported by the Canada Research Chair in Statistical Trial Design and funded by the Discovery Grant Program of the Natural Sciences and Engineering Research Council of Canada (RGPIN-2021-03366).

\subsection*{Financial disclosure}

Funding agreements ensure the authors’ independence in developing the methodology, designing the simulation study, interpreting the results, writing, and publishing the article.

\subsection*{Conflict of interest}

The authors declare no potential conflict of interests.

\subsection*{Data Availability Statement}

The files required to generate the data, run the simulations, and reproduce the results are available at \url{http://github.com/remiroazocar/Gcomp_indirect_comparisons_simstudy}.

\subsection*{Highlights}

\paragraph{What is already known?}

\begin{itemize}
\item Population adjustment methods such as matching-adjusted indirect comparison (MAIC) are increasingly used to compare marginal treatment effects where there are cross-trial differences in effect modifiers and limited patient-level data.

\item Current outcome regression-based alternatives, such as the conventional usage of simulated treatment comparison (STC), target a conditional treatment effect that is incompatible in the indirect comparison. Marginalization methods are required for compatible indirect treatment comparisons. 
\end{itemize}

\paragraph{What is new?}

\begin{itemize}
\item We present a marginalization method based on parametric G-computation that can be easily applied where the outcome regression is a generalized linear model. The method can accommodate a Bayesian statistical framework, which integrates the analysis into a probabilistic framework, typically required for health technology assessment.

\item We conduct a simulation study that provides proof-of-principle and benchmarks the performance of parametric G-computation against MAIC and the conventional use of STC. 
\end{itemize}

\paragraph{Potential impact for RSM readers outside the authors’ field}

\begin{itemize}
\item MAIC is the most widely used method for pairwise population-adjusted indirect comparisons. 

\item In the scenarios we considered, parametric G-computation achieves more precise and more accurate estimates than MAIC, particularly when covariate overlap is poor. It yields unbiased treatment effect estimates under correct model specification.

\item In addition, parametric G-computation provides lower empirical standard errors and mean square errors than the conventional approach to outcome regression, which, in addition, suffers from non-collapsibility bias. 

\item G-computation methods should be considered for population-adjusted indirect comparisons, particularly where there is limited covariate overlap, and this leads to extreme weights and small effective sample sizes in MAIC.
\end{itemize}

%\nocite{*}% Show all bib entries - both cited and uncited; comment this line to view only cited bib entries;
\bibliography{wileyNJD-AMA}

\begin{thebibliography}{100}
\providecommand \doibase [0]{http://dx.doi.org/}%

\bibitem{vreman2020decision}
Vreman RA, Naci H, Goettsch WG, et al. Decision making under uncertainty:
  comparing regulatory and health technology assessment reviews of medicines in
  the United States and Europe. {\it Clinical Pharmacology \& Therapeutics}
  2020\string; 108(2)\string: 350--357.

\bibitem{temple2000placebo}
Temple R, Ellenberg SS. Placebo-controlled trials and active-control trials in
  the evaluation of new treatments. Part 1: ethical and scientific issues. {\it
  Annals of internal medicine} 2000\string; 133(6)\string: 455--463.

\bibitem{paul2001fourth}
Paul JE, Trueman P. ‘Fourth hurdle reviews’, NICE, and database
  applications. {\it Pharmacoepidemiology and drug safety} 2001\string;
  10(5)\string: 429--438.

\bibitem{sutton2008use}
Sutton A, Ades A, Cooper N, Abrams K. Use of indirect and mixed treatment
  comparisons for technology assessment. {\it Pharmacoeconomics} 2008\string;
  26(9)\string: 753--767.

\bibitem{dias2013evidence}
Dias S, Sutton AJ, Ades A, Welton NJ. Evidence synthesis for decision making 2:
  a generalized linear modeling framework for pairwise and network
  meta-analysis of randomized controlled trials. {\it Medical Decision Making}
  2013\string; 33(5)\string: 607--617.

\bibitem{bucher1997results}
Bucher HC, Guyatt GH, Griffith LE, Walter SD. The results of direct and
  indirect treatment comparisons in meta-analysis of randomized controlled
  trials. {\it Journal of clinical epidemiology} 1997\string; 50(6)\string:
  683--691.

\bibitem{phillippo2018methods}
Phillippo DM, Ades AE, Dias S, Palmer S, Abrams KR, Welton NJ. Methods for
  population-adjusted indirect comparisons in health technology appraisal. {\it
  Medical Decision Making} 2018\string; 38(2)\string: 200--211.

\bibitem{signorovitch2010comparative}
Signorovitch JE, Wu EQ, Andrew PY, et al. Comparative effectiveness without
  head-to-head trials. {\it Pharmacoeconomics} 2010\string; 28(10)\string:
  935--945.

\bibitem{caro2010no}
Caro JJ, Ishak KJ. No head-to-head trial? Simulate the missing arms. {\it
  Pharmacoeconomics} 2010\string; 28(10)\string: 957--967.

\bibitem{miettinen1972standardization}
Miettinen OS. Standardization of risk ratios. {\it American Journal of
  Epidemiology} 1972\string; 96(6)\string: 383--388.

\bibitem{stuart2011use}
Stuart EA, Cole SR, Bradshaw CP, Leaf PJ. The use of propensity scores to
  assess the generalizability of results from randomized trials. {\it Journal
  of the Royal Statistical Society: Series A (Statistics in Society)}
  2011\string; 174(2)\string: 369--386.

\bibitem{phillippo2020multilevel}
Phillippo DM, Dias S, Ades A, et al. Multilevel network meta-regression for
  population-adjusted treatment comparisons. {\it Journal of the Royal
  Statistical Society: Series A (Statistics in Society)} 2020.

\bibitem{phillippo2019calibration}
Phillippo DM. {\it Calibration of treatment effects in network meta-analysis
  using individual patient data}. PhD thesis. University of Bristol, Bristol,
  UK;  2019.

\bibitem{phillippo2016nice}
Phillippo D, Ades T, Dias S, Palmer S, Abrams KR, Welton N. NICE DSU technical
  support document 18: methods for population-adjusted indirect comparisons in
  submissions to NICE.  2016.

\bibitem{baio2012bayesian}
Baio G. {\it Bayesian methods in health economics}.
\newblock CRC Press .
\newblock 2012.

\bibitem{claxton2005probabilistic}
Claxton K, Sculpher M, McCabe C, et al. Probabilistic sensitivity analysis for
  NICE technology assessment: not an optional extra. {\it Health economics}
  2005\string; 14(4)\string: 339--347.

\bibitem{remiro2020methods}
Remiro-Az{\'o}car A, Heath A, Baio G. Methods for Population Adjustment with
  Limited Access to Individual Patient Data: A Review and Simulation Study.
  {\it arXiv preprint arXiv:2004.14800} 2020.

\bibitem{cheng2019statistical}
Cheng D, Ayyagari R, Signorovitch J. The Statistical Performance of
  Matching-Adjusted Indirect Comparisons. {\it arXiv preprint arXiv:1910.06449}
  2019.

\bibitem{hatswell2020effects}
Hatswell AJ, Freemantle N, Baio G. The Effects of Model Misspecification in
  Unanchored Matching-Adjusted Indirect Comparison (MAIC): Results of a
  Simulation Study. {\it Value in Health} 2020.

\bibitem{phillippo2020assessing}
Phillippo DM, Dias S, Ades A, Welton NJ. Assessing the performance of
  population adjustment methods for anchored indirect comparisons: A simulation
  study. {\it Statistics in Medicine} 2020\string; 39(30)\string: 4885--4911.

\bibitem{stuart2010matching}
Stuart EA. Matching methods for causal inference: A review and a look forward.
  {\it Statistical science: a review journal of the Institute of Mathematical
  Statistics} 2010\string; 25(1)\string: 1.

\bibitem{jackson2020alternative}
Jackson D, Rhodes K, Ouwens M. Alternative weighting schemes when performing
  matching-adjusted indirect comparisons. {\it Research Synthesis Methods}
  2020.

\bibitem{van2011targeted}
Laan V.~dMJ, Rose S. {\it Targeted learning: causal inference for observational
  and experimental data}.
\newblock Springer Science \& Business Media .
\newblock 2011.

\bibitem{neugebauer2005prefer}
Neugebauer R, Laan v.~dM. Why prefer double robust estimators in causal
  inference?. {\it Journal of statistical planning and inference} 2005\string;
  129(1-2)\string: 405--426.

\bibitem{robins1992estimating}
Robins JM, Mark SD, Newey WK. Estimating exposure effects by modelling the
  expectation of exposure conditional on confounders. {\it Biometrics}
  1992\string: 479--495.

\bibitem{vo2021assessing}
Vo TT, Porcher R, Vansteelandt S. Assessing the impact of case-mix
  heterogeneity in individual participant data meta-analysis: Novel use of I 2
  statistic and prediction interval. {\it Research Methods in Medicine \&
  Health Sciences} 2021\string; 2(1)\string: 12--30.

\bibitem{phillippo2019population}
Phillippo DM, Dias S, Elsada A, Ades A, Welton NJ. Population Adjustment
  Methods for Indirect Comparisons: A Review of National Institute for Health
  and Care Excellence Technology Appraisals. {\it International journal of
  technology assessment in health care} 2019\string: 1--8.

\bibitem{hauck1998should}
Hauck WW, Anderson S, Marcus SM. Should we adjust for covariates in nonlinear
  regression analyses of randomized trials?. {\it Controlled clinical trials}
  1998\string; 19(3)\string: 249--256.

\bibitem{daniel2020making}
Daniel R, Zhang J, Farewell D. Making apples from oranges: Comparing
  noncollapsible effect estimators and their standard errors after adjustment
  for different covariate sets. {\it Biometrical Journal} 2020.

\bibitem{robins1986new}
Robins J. A new approach to causal inference in mortality studies with a
  sustained exposure period—application to control of the healthy worker
  survivor effect. {\it Mathematical modelling} 1986\string; 7(9-12)\string:
  1393--1512.

\bibitem{robins1987graphical}
Robins J. A graphical approach to the identification and estimation of causal
  parameters in mortality studies with sustained exposure periods. {\it Journal
  of chronic diseases} 1987\string; 40\string: 139S--161S.

\bibitem{moore2009covariate}
Moore KL, Laan v.~dMJ. Covariate adjustment in randomized trials with binary
  outcomes: targeted maximum likelihood estimation. {\it Statistics in
  medicine} 2009\string; 28(1)\string: 39--64.

\bibitem{austin2010absolute}
Austin PC. Absolute risk reductions, relative risks, relative risk reductions,
  and numbers needed to treat can be obtained from a logistic regression model.
  {\it Journal of clinical epidemiology} 2010\string; 63(1)\string: 2--6.

\bibitem{vo2019novel}
Vo TT, Porcher R, Chaimani A, Vansteelandt S. A novel approach for identifying
  and addressing case-mix heterogeneity in individual participant data
  meta-analysis. {\it Research synthesis methods} 2019\string; 10(4)\string:
  582--596.

\bibitem{remiro2021target}
Remiro-Az{\'o}car A. Target estimands for population-adjusted indirect
  comparisons. {\it arXiv preprint arXiv:2112.08023} 2021.

\bibitem{austin2011introduction}
Austin PC. An introduction to propensity score methods for reducing the effects
  of confounding in observational studies. {\it Multivariate behavioral
  research} 2011\string; 46(3)\string: 399--424.

\bibitem{hernan2020causal}
Hern{\'a}n MA, Robins JM. {\it Causal inference: what if}.
\newblock Boca Raton: Chapman \& Hall/CRC .
\newblock 2020.

\bibitem{remiro2021marginalization}
Remiro-Az{\'o}car A, Heath A, Baio G. The marginalization of
  regression-adjusted estimates is necessary for reimbursement decisions at the
  population level.

\bibitem{vanderweele2009concerning}
VanderWeele TJ. Concerning the consistency assumption in causal inference. {\it
  Epidemiology} 2009\string; 20(6)\string: 880--883.

\bibitem{rothman1980concepts}
Rothman KJ, Greenland S, Walker AM. Concepts of interaction. {\it American
  journal of epidemiology} 1980\string; 112(4)\string: 467--470.

\bibitem{manski2019meta}
Manski CF. Meta-analysis for medical decisions.  2019.

\bibitem{song2003validity}
Song F, Altman DG, Glenny AM, Deeks JJ. Validity of indirect comparison for
  estimating efficacy of competing interventions: empirical evidence from
  published meta-analyses. {\it Bmj} 2003\string; 326(7387)\string: 472.

\bibitem{remiro2020conflating}
Remiro-Az{\'o}car A, Heath A, Baio G. Conflating marginal and conditional
  treatment effects: Comments on'Assessing the performance of population
  adjustment methods for anchored indirect comparisons: A simulation study'.
  {\it arXiv preprint arXiv:2011.06334} 2020.

\bibitem{cole2010generalizing}
Cole SR, Stuart EA. Generalizing evidence from randomized clinical trials to
  target populations: the ACTG 320 trial. {\it American journal of
  epidemiology} 2010\string; 172(1)\string: 107--115.

\bibitem{remiro2020principled}
Remiro-Az{\'o}car A, Heath A, Baio G. Principled selection of effect modifiers:
  Comments on'Matching-adjusted indirect comparisons: Application to
  time-to-event data'. {\it arXiv preprint arXiv:2012.05127} 2020.

\bibitem{nelsen2007introduction}
Nelsen RB. {\it An introduction to copulas}.
\newblock Springer Science \& Business Media .
\newblock 2007.

\bibitem{sklar1959fonctions}
Sklar M. Fonctions de repartition an dimensions et leurs marges. {\it Publ.
  Inst. Statist. Univ. Paris} 1959\string; 8\string: 229--231.

\bibitem{kruschke2014doing}
Kruschke J. {\it Doing Bayesian data analysis: A tutorial with R, JAGS, and
  Stan}.
\newblock Academic Press .
\newblock 2014.

\bibitem{royston2004multiple}
Royston P. Multiple imputation of missing values. {\it The Stata Journal}
  2004\string; 4(3)\string: 227--241.

\bibitem{buuren2010mice}
Buuren Sv, Groothuis-Oudshoorn K. mice: Multivariate imputation by chained
  equations in R. {\it Journal of statistical software} 2010\string: 1--68.

\bibitem{ishak2015simulation}
Ishak KJ, Proskorovsky I, Benedict A. Simulation and matching-based approaches
  for indirect comparison of treatments. {\it Pharmacoeconomics} 2015\string;
  33(6)\string: 537--549.

\bibitem{petto2019alternative}
Petto H, Kadziola Z, Brnabic A, Saure D, Belger M. Alternative Weighting
  Approaches for Anchored Matching-Adjusted Indirect Comparisons via a Common
  Comparator. {\it Value in Health} 2019\string; 22(1)\string: 85--91.

\bibitem{belger2015inclusion}
Belger M, Brnabic A, Kadziola Z, Petto H, Faries D. Inclusion of multiple
  studies in matching adjusted indirect comparisons (MAIC). {\it Value in
  Health} 2015\string; 18(3)\string: A33.

\bibitem{grimm2019nivolumab}
Grimm SE, Armstrong N, Ramaekers BL, et al. Nivolumab for treating metastatic
  or unresectable urothelial cancer: an evidence review group perspective of a
  NICE single technology appraisal. {\it Pharmacoeconomics} 2019\string;
  37(5)\string: 655--667.

\bibitem{ren2019pembrolizumab}
Ren S, Squires H, Hock E, Kaltenthaler E, Rawdin A, Alifrangis C. Pembrolizumab
  for Locally Advanced or Metastatic Urothelial Cancer Where Cisplatin is
  Unsuitable: An Evidence Review Group Perspective of a NICE Single Technology
  Appraisal. {\it PharmacoEconomics} 2019\string; 37(9)\string: 1073--1080.

\bibitem{janes2010quantifying}
Janes H, Dominici F, Zeger S. On quantifying the magnitude of confounding. {\it
  Biostatistics} 2010\string; 11(3)\string: 572--582.

\bibitem{greenland1987interpretation}
Greenland S. Interpretation and choice of effect measures in epidemiologic
  analyses. {\it American journal of epidemiology} 1987\string; 125(5)\string:
  761--768.

\bibitem{greenland1999confounding}
Greenland S, Robins JM, Pearl J. Confounding and collapsibility in causal
  inference. {\it Statistical science} 1999\string: 29--46.

\bibitem{austin2014use}
Austin PC. The use of propensity score methods with survival or time-to-event
  outcomes: reporting measures of effect similar to those used in randomized
  experiments. {\it Statistics in medicine} 2014\string; 33(7)\string:
  1242--1258.

\bibitem{keil2018bayesian}
Keil AP, Daza EJ, Engel SM, Buckley JP, Edwards JK. A Bayesian approach to the
  g-formula. {\it Statistical methods in medical research} 2018\string;
  27(10)\string: 3183--3204.

\bibitem{vansteelandt2011invited}
Vansteelandt S, Keiding N. Invited commentary: G-computation--lost in
  translation?. {\it American journal of epidemiology} 2011\string;
  173(7)\string: 739--742.

\bibitem{imbens2015causal}
Imbens GW, Rubin DB. {\it Causal inference in statistics, social, and
  biomedical sciences}.
\newblock Cambridge University Press .
\newblock 2015.

\bibitem{phillippo2021target}
Phillippo DM, Dias S, Ades AE, Welton NJ. Target estimands for efficient
  decision making: Response to comments on “Assessing the performance of
  population adjustment methods for anchored indirect comparisons: A simulation
  study”. {\it Statistics in Medicine} 2021\string; 40(11)\string:
  2759--2763.

\bibitem{zhang2016new}
Zhang Z, Nie L, Soon G, Hu Z. New methods for treatment effect calibration,
  with applications to non-inferiority trials. {\it Biometrics} 2016\string;
  72(1)\string: 20--29.

\bibitem{efron1986bootstrap}
Efron B, Tibshirani R. Bootstrap methods for standard errors, confidence
  intervals, and other measures of statistical accuracy. {\it Statistical
  science} 1986\string: 54--75.

\bibitem{varadhan2016cross}
Varadhan R, Henderson NC, Weiss CO. Cross-design synthesis for extending the
  applicability of trial evidence when treatment effect is heterogeneous: Part
  I. Methodology. {\it Communications in Statistics: Case Studies, Data
  Analysis and Applications} 2016\string; 2(3-4)\string: 112--126.

\bibitem{rubin1987logit}
Rubin DB, Schenker N. Logit-based interval estimation for binomial data using
  the Jeffreys prior. {\it Sociological methodology} 1987\string: 131--144.

\bibitem{aalen1997markov}
Aalen OO, Farewell VT, De~Angelis D, Day NE, N{\"o}el~Gill O. A Markov model
  for HIV disease progression including the effect of HIV diagnosis and
  treatment: application to AIDS prediction in England and Wales. {\it
  Statistics in medicine} 1997\string; 16(19)\string: 2191--2210.

\bibitem{keil2014autism}
Keil AP, Daniels JL, Hertz-Picciotto I. Autism spectrum disorder, flea and tick
  medication, and adjustments for exposure misclassification: the CHARGE
  (CHildhood Autism Risks from Genetics and Environment) case--control study.
  {\it Environmental Health} 2014\string; 13(1)\string: 1--10.

\bibitem{josefsson2021bayesian}
Josefsson M, Daniels MJ. Bayesian semi-parametric G-computation for causal
  inference in a cohort study with MNAR dropout and death. {\it Journal of the
  Royal Statistical Society: Series C (Applied Statistics)} 2021\string;
  70(2)\string: 398--414.

\bibitem{rubin1978bayesian}
Rubin DB. Bayesian inference for causal effects: The role of randomization.
  {\it The Annals of statistics} 1978\string: 34--58.

\bibitem{saarela2015predictive}
Saarela O, Arjas E, Stephens DA, Moodie EE. Predictive Bayesian inference and
  dynamic treatment regimes. {\it Biometrical Journal} 2015\string;
  57(6)\string: 941--958.

\bibitem{lunn2012bugs}
Lunn D, Jackson C, Best N, Thomas A, Spiegelhalter D. {\it The BUGS book: A
  practical introduction to Bayesian analysis}.
\newblock CRC press .
\newblock 2012.

\bibitem{plummer2003jags}
Plummer M, others . JAGS: A program for analysis of Bayesian graphical models
  using Gibbs sampling. In: . 124. Vienna, Austria. ; 2003\string: 1--10.

\bibitem{carpenter2017stan}
Carpenter B, Gelman A, Hoffman MD, et al. Stan: A probabilistic programming
  language. {\it Journal of statistical software} 2017\string; 76(1).

\bibitem{rue2009approximate}
Rue H, Martino S, Chopin N. Approximate Bayesian inference for latent Gaussian
  models by using integrated nested Laplace approximations. {\it Journal of the
  royal statistical society: Series b (statistical methodology)} 2009\string;
  71(2)\string: 319--392.

\bibitem{guyot2012enhanced}
Guyot P, Ades A, Ouwens MJ, Welton NJ. Enhanced secondary analysis of survival
  data: reconstructing the data from published Kaplan-Meier survival curves.
  {\it BMC medical research methodology} 2012\string; 12(1)\string: 9.

\bibitem{morris2019using}
Morris TP, White IR, Crowther MJ. Using simulation studies to evaluate
  statistical methods. {\it Statistics in medicine} 2019\string; 38(11)\string:
  2074--2102.

\bibitem{team2013r}
Team RC, others . R: A language and environment for statistical computing.
  2013.

\bibitem{ripley2009stochastic}
Ripley BD. {\it Stochastic simulation}. 316.
\newblock John Wiley \& Sons .
\newblock 2009.

\bibitem{skipka2016methodological}
Skipka G, Wieseler B, Kaiser T, et al. Methodological approach to determine
  minor, considerable, and major treatment effects in the early benefit
  assessment of new drugs. {\it Biometrical Journal} 2016\string; 58(1)\string:
  43--58.

\bibitem{stanley2007design}
Stanley K. Design of randomized controlled trials. {\it Circulation}
  2007\string; 115(9)\string: 1164--1169.

\bibitem{cohen2013statistical}
Cohen J. {\it Statistical power analysis for the behavioral sciences}.
\newblock Academic press .
\newblock 2013.

\bibitem{vittinghoff2007relaxing}
Vittinghoff E, McCulloch CE. Relaxing the rule of ten events per variable in
  logistic and Cox regression. {\it American journal of epidemiology}
  2007\string; 165(6)\string: 710--718.

\bibitem{neyman1934two}
Neyman J. On the two different aspects of the representative method: the method
  of stratified sampling and the method of purposive selection. {\it Journal of
  the Royal Statistical Society} 1934\string; 97(4)\string: 558--625.

\bibitem{annesi1989efficiency}
Annesi I, Moreau T, Lellouch J. Efficiency of the logistic regression and Cox
  proportional hazards models in longitudinal studies. {\it Statistics in
  medicine} 1989\string; 8(12)\string: 1515--1521.

\bibitem{ho2007matching}
Ho DE, Imai K, King G, Stuart EA. Matching as nonparametric preprocessing for
  reducing model dependence in parametric causal inference. {\it Political
  analysis} 2007\string; 15(3)\string: 199--236.

\bibitem{phillippo2020equivalence}
Phillippo DM, Dias S, Ades A, Welton NJ. Equivalence of entropy balancing and
  the method of moments for matching-adjusted indirect comparison. {\it
  Research Synthesis Methods} 2020.

\bibitem{dehejia1999causal}
Dehejia RH, Wahba S. Causal effects in nonexperimental studies: Reevaluating
  the evaluation of training programs. {\it Journal of the American statistical
  Association} 1999\string; 94(448)\string: 1053--1062.

\bibitem{waernbaum2010propensity}
Waernbaum I. Propensity score model specification for estimation of average
  treatment effects. {\it Journal of Statistical Planning and Inference}
  2010\string; 140(7)\string: 1948--1956.

\bibitem{westreich2019target}
Westreich D, Edwards JK, Lesko CR, Cole SR, Stuart EA. Target validity and the
  hierarchy of study designs. {\it American journal of epidemiology}
  2019\string; 188(2)\string: 438--443.

\bibitem{imai2008misunderstandings}
Imai K, King G, Stuart EA. Misunderstandings between experimentalists and
  observationalists about causal inference. {\it Journal of the royal
  statistical society: series A (statistics in society)} 2008\string;
  171(2)\string: 481--502.

\bibitem{lunceford2004stratification}
Lunceford JK, Davidian M. Stratification and weighting via the propensity score
  in estimation of causal treatment effects: a comparative study. {\it
  Statistics in medicine} 2004\string; 23(19)\string: 2937--2960.

\bibitem{van2003unified}
Laan V.~dMJ, Laan M, Robins JM. {\it Unified methods for censored longitudinal
  data and causality}.
\newblock Springer Science \& Business Media .
\newblock 2003.

\bibitem{bang2005doubly}
Bang H, Robins JM. Doubly robust estimation in missing data and causal
  inference models. {\it Biometrics} 2005\string; 61(4)\string: 962--973.

\bibitem{kang2007demystifying}
Kang JD, Schafer JL, others . Demystifying double robustness: A comparison of
  alternative strategies for estimating a population mean from incomplete data.
  {\it Statistical science} 2007\string; 22(4)\string: 523--539.

\bibitem{ishak2015simulated}
Ishak K, Rael M, Phatak H, Masseria C, Lanitis T. Simulated treatment
  comparison of time-to-event (and other non-linear) outcomes. {\it Value in
  Health} 2015\string; 18(7)\string: A719.

\bibitem{bartlett2018covariate}
Bartlett JW. Covariate adjustment and estimation of mean response in randomised
  trials. {\it Pharmaceutical statistics} 2018\string; 17(5)\string: 648--666.

\bibitem{qu2015estimation}
Qu Y, Luo J. Estimation of group means when adjusting for covariates in
  generalized linear models. {\it Pharmaceutical statistics} 2015\string;
  14(1)\string: 56--62.

\bibitem{lane1982analysis}
Lane PW, Nelder JA. Analysis of covariance and standardization as instances of
  prediction. {\it Biometrics} 1982\string: 613--621.

\bibitem{chan2014increasing}
Chan AW, Song F, Vickers A, et al. Increasing value and reducing waste:
  addressing inaccessible research. {\it The Lancet} 2014\string;
  383(9913)\string: 257--266.

\bibitem{nowok2016synthpop}
Nowok B, Raab GM, Dibben C, others . synthpop: Bespoke creation of synthetic
  data in R. {\it J Stat Softw} 2016\string; 74(11)\string: 1--26.

\bibitem{bonofigliorecovery}
Bonofiglio F, Schumacher M, Binder H. Recovery of original individual person
  data (IPD) inferences from empirical IPD summaries only: Applications to
  distributed computing under disclosure constraints. {\it Statistics in
  Medicine}.

\bibitem{faria2015nice}
Faria R, Hernandez~Alava M, Manca A, Wailoo A. NICE DSU technical support
  document 17: the use of observational data to inform estimates of treatment
  effectiveness for technology appraisal: methods for comparative individual
  patient data. {\it Sheffield: NICE Decision Support Unit} 2015.

\bibitem{robins2009estimation}
Robins JM, Hern{\'a}n MA. Estimation of the causal effects of time-varying
  exposures. {\it Longitudinal data analysis} 2009\string; 553\string: 599.

\end{thebibliography}

\clearpage

\begin{center}
\begin{table*}[t]%
\caption{
The implementation of the methodologies compared in the simulation study. See Appendix A of the Supplementary Material for a more a detailed description of the methods and their specific settings. For all methods, the marginal log-odds ratio for $B$ vs.~$C$ is estimated directly from the event counts, and its standard error is computed using the delta method. The marginal log-odds ratio estimate for $A$ vs.~$B$ and its standard error are obtained by combining the within-study point estimates. Wald-type 95\% interval estimates are constructed for the marginal $A$ vs.~$B$ treatment effect using normal distributions.\label{tab1}}
\centering
\begin{tabular*}{500pt}
{p{7cm} p{} p{10cm}}
\toprule
\textbf{Method} & \multicolumn{2}{l}{\textbf{Details}}  \\
\midrule
Matching-adjusted indirect comparison (MAIC) & \multicolumn{2}{p{9cm}}{\raggedright 
\begin{itemize}[noitemsep, topsep=0pt]
    \item Method of moments formulation presented by Signorovitch et al.\cite{signorovitch2010comparative, phillippo2016nice, remiro2020methods, phillippo2020equivalence} 
    \item Effect modifier means balanced for active treatment and control arms combined
    \item Standard errors for the $A$ vs.~$C$ treatment effect computed by resampling via the ordinary non-parametric bootstrap with replacement (1,000 resamples)
\end{itemize}
} \\
Simulated treatment comparison (STC) &
\multicolumn{2}{p{9cm}}{\raggedright 
\begin{itemize}[noitemsep, topsep=0pt]
    \item Conventional version as described by HTA guidance and recommendations\cite{phillippo2016nice}
    \item Correctly specified covariate-adjusted logistic regression fitted to the $AC$ IPD using maximum-likelihood estimation
    \item Effect modifiers are centered at their mean $BC$ values 
\end{itemize}    
}\\
Maximum-likelihood parametric G-computation & \multicolumn{2}{p{9cm}}
{
\begin{itemize}[noitemsep, topsep=0pt]
    \item Correctly specified covariate-adjusted logistic regression fitted to the $AC$ IPD using maximum-likelihood estimation
    \item $BC$ covariate simulation from Gaussian copula using normally-distributed marginals with the $BC$ means and standard deviations, and the pairwise linear correlations of the $AC$ IPD ($N^*=1000$)
    \item Outcomes in $BC$ predicted by plugging the simulated covariates into the maximum-likelihood fit
    \item Standard errors for the $A$ vs.~$C$ treatment effect computed by resampling using the ordinary non-parametric bootstrap with replacement (1,000 resamples)
\end{itemize}
} \\
Bayesian parametric G-computation & 
\multicolumn{2}{p{9cm}}{\raggedright 
\begin{itemize}[noitemsep, topsep=0pt]
\item Correctly specified covariate-adjusted logistic regression fitted to the $AC$ IPD using Markov chain Monte Carlo (weakly informative priors, 2 chains with 4,000 total iterations each, including 2,000 warmup iterations each) 
\item $BC$ covariate simulation from Gaussian copula using normally-distributed marginals with the $BC$ means and standard deviations, and the pairwise linear correlations of the $AC$ IPD ($N^*=1000$)
\item Outcomes in $BC$ predicted by drawing from their posterior predictive distribution
\end{itemize}
} \\
\bottomrule
% \begin{tablenotes}%%[341pt]
% \item Source: Example for table source text.
% \item[1] Example for a first table footnote.
% \item[2] Example for a second table footnote.
% \end{tablenotes}
\end{tabular*}
\end{table*}
\end{center}

\end{document}